\documentclass[11pt,a4paper]{article}
\pdfoutput=1

\usepackage{amsmath}
\usepackage{color}
\usepackage{amsfonts}
\usepackage{amssymb}
\usepackage{graphicx}
\usepackage{geometry}
\usepackage{amssymb,epsfig,subfigure}
\usepackage{hyperref}
\usepackage[font=footnotesize]{caption}
\usepackage{comment}

\usepackage{dsfont}
\usepackage{diagbox}

\usepackage{euscript}     
\usepackage{braket}
\usepackage{starfont}
\usepackage{color,soul}         
\usepackage{tensor}        
\usepackage{amsthm}

\usepackage{slashed}
\usepackage{leftidx}
\usepackage{subfigure}
\usepackage{bbm}
\usepackage{authblk}

\usepackage[numbers,sort&compress]{natbib}  
\usepackage{float}

\usepackage{ytableau}

\makeatletter
\renewcommand\section{\@startsection {section}{1}{\z@}%
                                 {-3.5ex \@plus -1ex \@minus -.2ex}
                                   {2.3ex \@plus.2ex}%
                                   {\normalfont\large\bfseries}}
\renewcommand\subsection{\@startsection{subsection}{2}{\z@}%
                                   {-3.25ex\@plus -1ex \@minus -.2ex}%
                                     {1.5ex \@plus .2ex}%
                                     {\normalfont\bfseries}}
\renewcommand\subsubsection{\@startsection{subsubsection}{3}{\z@}%
                                   {-3.25ex\@plus -1ex \@minus -.2ex}%
                                     {1.5ex \@plus .2ex}%
                                     {\normalfont\itshape}}
\makeatother

\def\pplogo{\vbox{\kern-\headheight\kern -29pt
\halign{##&##\hfil\cr&{\ppnumber}\cr\rule{0pt}{2.5ex}&\ppdate\cr}}}
\makeatletter
\def\ps@firstpage{\ps@empty \def\@oddhead{\hss\pplogo}%
  \let\@evenhead\@oddhead 
}
\def\maketitle{\par
 \begingroup
 \def\thefootnote{\fnsymbol{footnote}}
 \def\@makefnmark{\hbox{$^{\@thefnmark}$\hss}}
 \if@twocolumn
 \twocolumn[\@maketitle]
 \else \newpage
 \global\@topnum\z@ \@maketitle \fi\thispagestyle{firstpage}\@thanks
 \endgroup
 \setcounter{footnote}{0}
 \let\maketitle\relax
 \let\@maketitle\relax
 \gdef\@thanks{}\gdef\@author{}\gdef\@title{}\let\thanks\relax}
\makeatother

\numberwithin{equation}{section}

\newcommand\eea{\end{eqnarray}}
\newcommand\bea{\begin{eqnarray}}

\def\la{\langle}
\def\ra{\rangle}
\def\beq{\begin{equation}}
\def\eeq{\end{equation}}

\def\la{\langle }
\def\ra{\rangle }
\newcommand{\be}{\begin{equation}}
\newcommand{\ee}{\end{equation}}
\newcommand{\ba}{\begin{align}}
\newcommand{\ea}{\end{align}}
\newcommand{\bg}{\begin{gather}}
\newcommand{\eg}{\end{gather}}
\newcommand{\bseq}{\begin{subequations}}
\newcommand{\eseq}{\end{subequations}}

\newcommand{\Tr}{{\rm Tr}}

\renewcommand{\t}{\tilde}

\newcommand{\tr}{{\rm tr}}

\begin{document}
\setcounter{page}0
\def\ppnumber{\vbox{\baselineskip14pt
}}
\def\ppdate{
} \date{}

\author[1]{Horacio Casini}
\author[2]{Eduardo Test\'e}
\author[1]{Gonzalo Torroba}
\affil[1]{Centro At\'omico Bariloche and CONICET, S.C. de Bariloche, R\'io Negro, R8402AGP, Argentina}
\affil[2]{Department of Physics, University of California, Santa Barbara, CA 93106, USA}
\setcounter{Maxaffil}{0}
\renewcommand\Affilfont{\itshape\small}


\bigskip
\title{\bf  Mutual information superadditivity \\ and unitarity bounds  \vskip 0.5cm}
\maketitle

\begin{abstract}
We derive the property of strong superadditivity of mutual information arising from the Markov property of the vacuum state in a conformal field theory and strong subadditivity of  entanglement entropy. We show this inequality encodes unitarity bounds for different types of fields. 
These unitarity bounds are precisely the ones that saturate for free fields. This has a natural explanation in terms of the possibility of localizing algebras on null surfaces. A particular continuity property of mutual information characterizes free fields from the entropic point of view. We derive a general formula for the leading long distance term of the mutual information for regions of arbitrary shape which involves the modular flow of these regions. We obtain the general form of this leading term for two spheres with arbitrary orientations in spacetime, and for primary fields of any tensor representation.  
For free fields we further obtain the explicit form of the leading term for arbitrary regions with boundaries on null cones.  
\end{abstract}
\bigskip

\newpage

\tableofcontents

\vskip 1cm

\section{Introduction}\label{sec:intro}

Entanglement entropy (EE) gives an unconventional description of quantum field theory (QFT) in terms of a statistical measure of the vacuum fluctuations in regions of space. It is a natural objective to understand if there is a full universal description of QFT by means of the EE. Several connections of EE with more conventional quantities have been understood. An important one relates to large distance entanglement.
  
Let us consider the mutual information between two well separated regions $A$, $B$ in vacuum.
 This is defined as the combination of entropies
 \be
 I(A,B)=S(A)+S(B)-S(A\cup B)\,,
 \ee
and has a meaning as a measure of the total amount of correlations between the regions.
 In a conformal field theory (CFT), and when the lowest dimension operator is a scalar of dimension $\Delta$, and the separation $L$ between the regions is much larger than their sizes, we have  
\be
I(A,B)\sim \frac{C(A)C(B)}{L^{4\Delta}}\,. \label{ilarge}
\ee
 This limit was analyzed  in \cite{Cardy.esferaslejanas} using a OPE expansion of the twist operators and further refined in \cite{agon2016quantum}, where the coefficients $C(A)$ were computed for the case of spherical regions. See \cite{Chen:2013kpa,Headrick:2010zt,Calabrese:2009ez,Calabrese:2010he,Rajabpour:2011pt} for previous $d=2$ analysis. In more generality, when the lowest dimension operator is not scalar some tensorial structures appear in the coefficients. See \cite{long2016co,chen2017mutual}. Subleading terms have also been computed \cite{Agon:2015twa,chen2017mutual,Chen:2016mya}, and some cases beyond the sphere are also known \cite{Schnitzer:2014zva}.
As a result of this analysis it is understood that the power series expansion of the mutual information for large distances contains the complete information of the spectrum of conformal dimensions of the CFT.

One of the main results of this work is a new superadditive inequality for the mutual information in CFTs, derived from the Markov property on the null cone and strong subadditivity of the entropy.
We will see that this provides another entry for the relationship between entropic quantities and general properties of QFTs. A basic nonperturbative result in unitary field theories is the existence of unitarity bounds of the form $\Delta \ge \Delta_*$ for the dimensions $\Delta$ of conformal primary operators. A violation of the bound implies the existence of certain negative norm states in the Hilbert space. These bounds also constrain the structure of general QFTs that start from ultraviolet conformal fixed points. We will show how mutual information superadditivity gives rise to certain unitarity bounds. Strong subadditivity contains information on the ``unitarity'' of the theory, meaning here the positive definite scalar product in the Hilbert space. However, its relation to the more standard manifestations of unitarity, like reflection positivity, have remained quite obscure. The relation  found in this paper of strong subadditivity and unitarity bounds provides a special instance in which this connection becomes more direct.\footnote{We also recall that strong subadditivity and the Markov property have been especially useful in the context of the proof of the irreversibility of the renormalization group flow in QFT \cite{Casini:2012ei, casini2017modular, casini2017markov, casini2018all}. }

An important aspect in our approach involves going beyond purely spatial regions for the mutual information, allowing for null deformations of their boundaries; see figs. \ref{hola}-\ref{deform} below. We find that saturation of superadditivity implies a certain geometric continuity of the mutual information that can only hold for free fields. In fact, this continuity actually defines what is a free field in entropic terms. Because of this, only the unitarity bounds for fields that have the tensor structure corresponding to conformal primary free fields are reproduced by the entropic inequalities.
It remains to understand if and how the general unitarity bounds can be derived from the mutual information.

In this paper we also further develop tools for computing the leading term of the mutual information in the long distance limit. We give a formula computing this term depending on the modular flows of the two regions. This generalizes the calculations in  \cite{Cardy.esferaslejanas,agon2016quantum}. With the help of this formula we give the explicit result for two spheres with arbitrary orientations in spacetime and for the contribution of primary fields in any tensor representation.  For spatial spheres we will show that the leading term has a remarkably simple form,
\be
I(A,B)\sim c(\Delta)\,\textrm{dim}({\cal R})\, \left(\frac{R_A R_B}{L^2}\right)^{2\Delta}\,, 
\ee 
where the coefficient depends only on the scaling dimension $\Delta$ and the dimension of the Lorentz representation ${\cal R}$ of the lowest dimensional primary field. For boosted spheres the mutual information depends on the lowest weights of ${\cal R}$, and is given below in section \ref{subsec:arbitrary}.
The result can be further extended for free fields to regions with arbitrary boundaries on the null cone.

The plan of the paper is as follows. In section \ref{sec:ssa} we describe the strong superadditivity of mutual information, and its application to the case of spheres in a CFT. We also show how the saturation of this inequality in the long distance limit can only happen for free fields. In section \ref{replica} we derive a formula for the coefficients of the mutual information in the long distance limit valid for any shapes for two regions.  In section \ref{examples} after computing these coefficients explicitly in several cases of interest, we derive the general formula for the case of two spheres with arbitrary orientations in spacetime and conformal primaries in general Lorentz representations. The formula for the coefficients based on modular flows allows us to prove the saturation of the inequalities in the free case from a different perspective in section \ref{sec:consec}. This result will allow us to give the explicit form of the coefficients for free fields and for any region in the light cone. 
In this same section we also compare in detail the entropic bounds with unitarity bounds. We end in section \ref{sec:concl} with a brief discussion and conclusions.  Two appendices contain additional technical details used in the main part of the paper.

\section{Mutual information superadditivity and the long distance expansion}\label{sec:ssa}

In this section we derive the strong superadditivity property of mutual information on the null cone and use it to obtain bounds on the 
 scaling behaviour of the mutual information at large distances. These bounds saturate for free fields.

\subsection{Strong superadditivity on the null cone}\label{subsec:ssa1}

For a CFT, the entanglement entropy of the vacuum in regions having a boundary on the null cone satisfies the Markov property. This is the saturation of strong subadditivity \cite{casini2017modular}. Calling $A_1$, $A_2$ to the two regions with boundaries on the null cone, we have for the entropies
\be
 S(A_1)+S(A_2)=S(A_1\cap A_2)+S(A_1\cup A_2)\,.\label{fgh}
\ee

\begin{figure}[t]
\begin{center}
\includegraphics[scale=0.13]{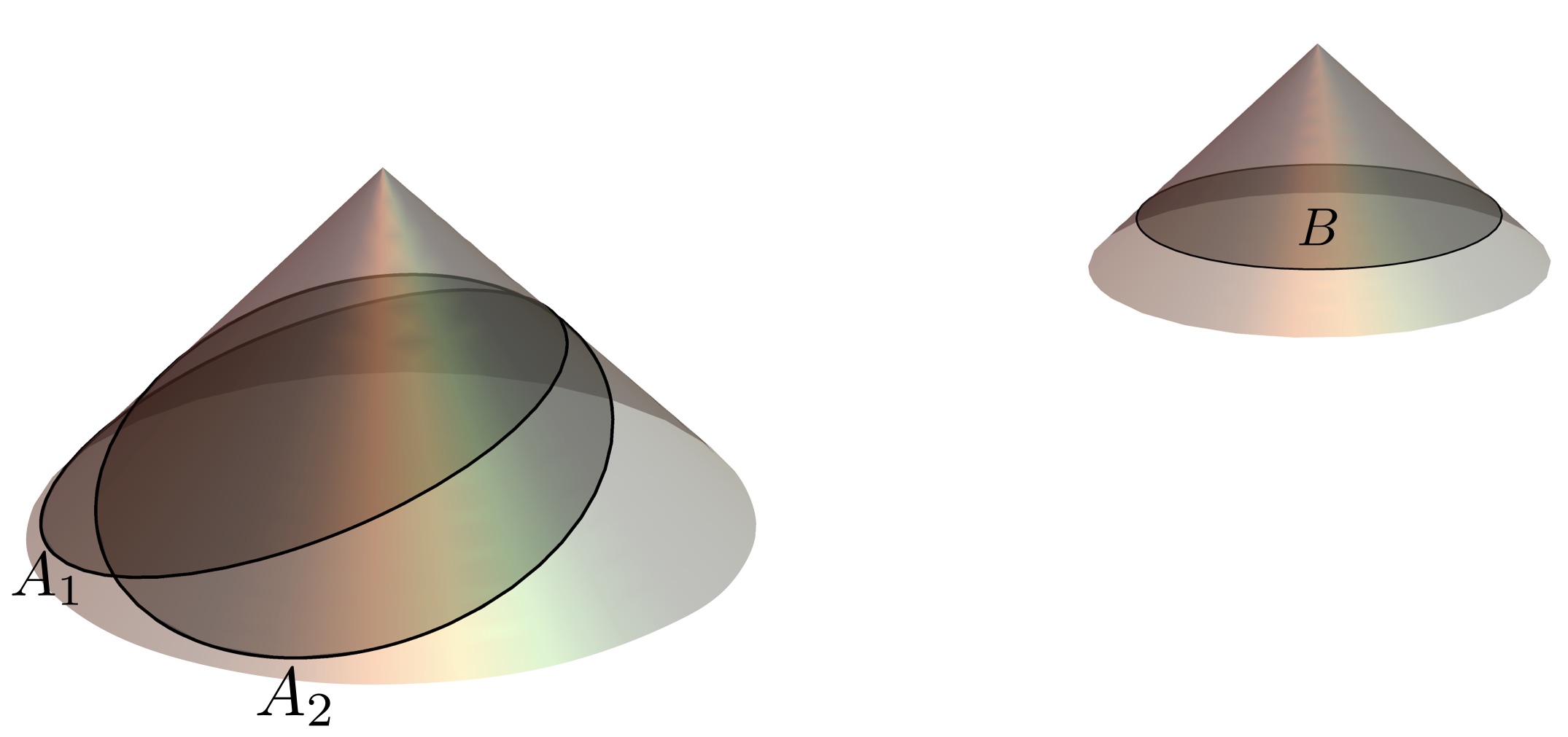}
\caption{Two regions $A_1, A_2$ with boundary on a null cone and another spatially separated region $B$.} \label{hola}
\end{center}
\end{figure}

Given two arbitrary space-like separated regions  $A$, $B$ we can compute the mutual information
\begin{equation}
I(A,B) = S(A) + S(B) - S(A \cup B)\,.
\end{equation}
 This is a well defined quantity in any type of operator algebras, including the type $III_1$ von Neumann algebras encountered in QFT. 
Taking two regions $A_1,A_2$ with boundaries on the null cone as above, and another spatially separated region $B$ (see figure \ref{hola}), we can compute
\bea
&&I(A_1\cap A_2,B)+I(A_1\cup A_2,B)-I(A_1,B)-I(A_2,B)\\
&& \hspace{3cm}=S(A_1\cup B)+S(A_2\cup B)-S((A_1\cap A_2)\cup B)-S((A_1\cup A_2)\cup B) \nonumber \,,
\eea
where we used (\ref{fgh}). The combination of entropies in the right hand side is always non-negative by the strong subadditivity property of entropy. Therefore, we obtain the strong superadditivity for the mutual information,
\be\label{eq:mutualSSA}
I(A_1\cap A_2,B)+I(A_1\cup A_2,B)-I(A_1,B)-I(A_2,B) \ge 0\,,
\ee
valid when one of the entries in the mutual information satisfies the Markov equation (\ref{fgh}). In the rest of the work we will study implications of this inequality, and its infinitesimal form, for CFTs.

\subsection{Two spheres and conformal invariance}\label{subsec:confinv}

In a conformal field theory the mutual information is conformally invariant. For the case of two spheres this is particularly powerful since the sphere is completely determined by the two time-like separated points at the tips of the causal domain of dependence. 
If we take the mutual information between two spheres, four points determine the configuration space of the mutual information (see \cite{long2016co}). Conformal symmetry implies that the mutual information must be a function of the two possible independent cross ratios between these four points \cite{ginsparg1988applied}. Moreover, it can be expanded in terms of conformal blocks, as we discuss in more detail in App. \ref{app:blocks}.

\begin{figure}[ht]
\begin{center}
\includegraphics[scale=0.15]{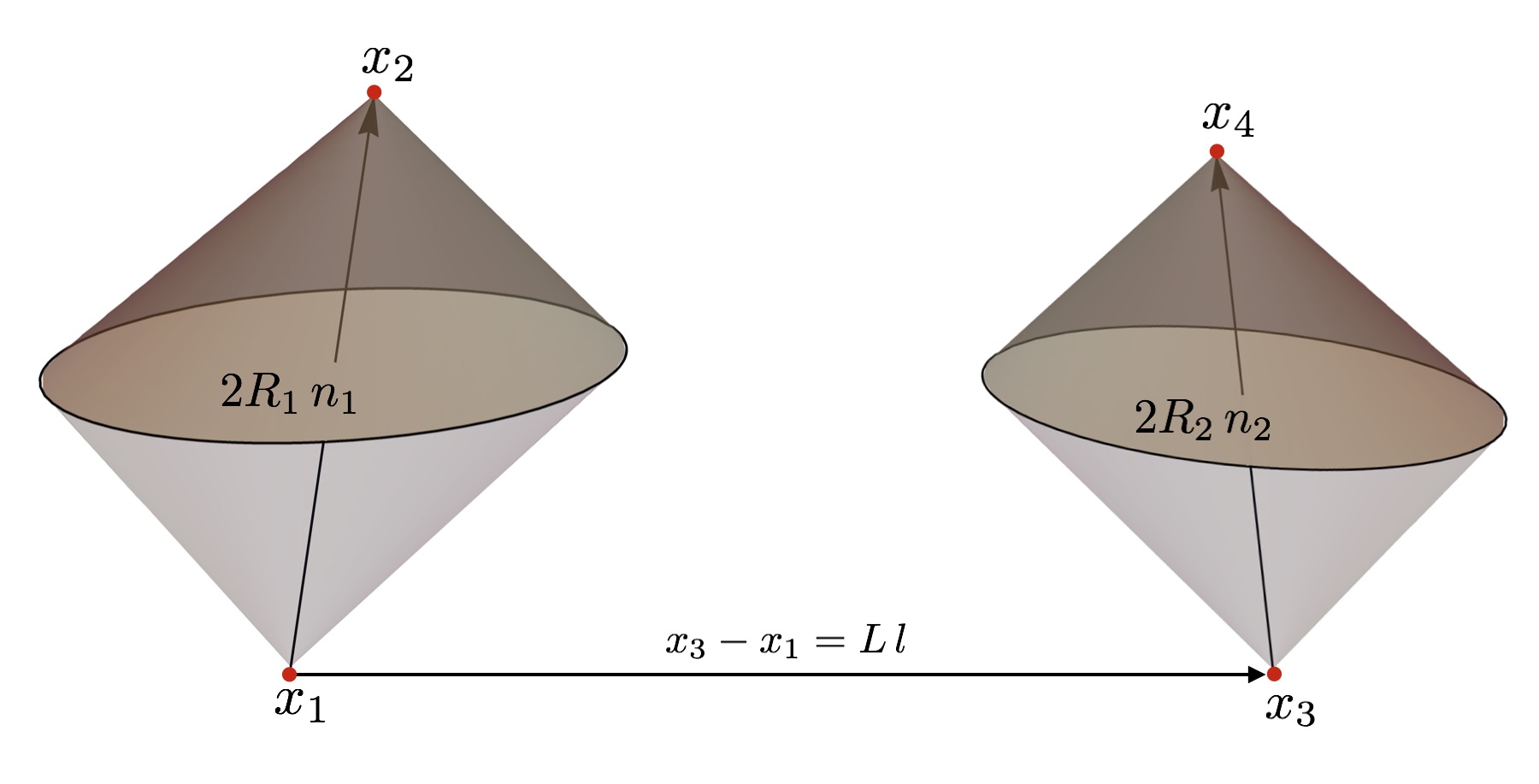}
\caption{Two spheres determined by four points $x_i$ located at the tips of their causal domains. In a CFT the mutual information between these two spheres is conformally invariant and depends solely on the two independent cross-ratios formed by $x_i$ in any dimension.} \label{two_spheres}
\end{center}
\end{figure}

Let us call the past and future tips of the causal diamond of the first sphere $x_1$ and $x_2$ respectively, and call $x_3$, $x_4$, the past and future tips corresponding to  the second sphere. We write 
\be\label{eq:n1n2}
x_2-x_1=2 R_1 n_1\;,\;x_4-x_3=2 R_2 n_2
\ee 
with $n_1$, $n_2$ two future directed unit time-like vectors ($n_1^2=n_2^2=-1$), and $R_1, R_2$ the radii of the spheres. We define the distance $L$ by 
\be\label{eq:l}
x_3-x_1=L \, l\,,
\ee 
with $l$ a unit spacelike vector, see Fig.~\ref{two_spheres}.

For our purpose, it will be convenient to choose  the following two independent cross ratios  
\bea
\chi_1 = \frac{|x_1-x_2||x_3-x_4|}{|x_1-x_3||x_2-x_4|} &=&  \frac{4 R_1  R_2}{L |2 R_1 n_1- L l-2 R_2 n_2|}\,,\label{chi1} \\
\chi_2 = \frac{|x_1-x_2||x_3-x_4|}{|x_1-x_4||x_2-x_3|} &=& \frac{4 R_1 R_2}{|2 R_1 n_1-L l| |2 R_2 n_2+L l|}  \,, \label{chi2}
\eea
where the norm is computed with the Minkowski signature $(- + \ldots +)$. The relation between $(\chi_1, \chi_2)$ and the conventional cross-ratios $(u, v)$ is
 \be\label{eq:uv0}
u = \frac{x_{12}^2 x_{34}^2}{x_{13}^2 x_{24}^2}= \chi_1^2\;,\; v= \frac{x_{14}^2 x_{23}^2}{x_{13}^2 x_{24}^2}= \frac{\chi_1^2}{\chi_2^2}\,.
\ee
For spatially separated spheres  $\chi_1\in (0,1)$, $\chi_2\in (0,\infty)$ and $\chi_2>\chi_1$. Both  cross ratios increase under inclusion of the spheres.   $I(\chi_1,\chi_2)$ is a positive monotonically increasing function under inclusion. 

We will be interested in the limit of long distance between the spheres. We will see that this limit has information on the conformal dimensions and spins of operators. Expanding for $L\gg R_1,R_2$ we have
\bea
\chi_1  &=& \frac{4 R_1 R_2}{L^2} + \frac{8 R_1 R_2}{L^2}\ l \cdot \left( n_1 \frac{ R_1}{L} -n_2 \frac{ R_2}{L}  \right)   + \left( \frac{4 R_1 R_2}{L^2} \right)^2 \left( (n_1 \cdot n_2) - 1\right) \\ && +8 R_1 R_2 \frac{(R_1+R_2)^2}{L^4}+ \frac{6 R_1 R_2}{L^2} \ \left( l\cdot \left( n_2 \frac{2 R_2}{L} - n_1 \frac{2 R_1}{L} \right)\right)^2       + {\cal O}((R/L)^5)\ , \nonumber \\
\chi_2  &=& \frac{4 R_1 R_2}{L^2} +\frac{8 R_1 R_2}{L^2}\ l \cdot \left( n_1 \frac{ R_1}{L} -n_2 \frac{ R_2}{L}  \right)  + \left( \frac{4 R_1 R_2}{L^2} \right)^2 \left( 2 (l \cdot n_1) (l \cdot n_2) - 1\right) \nonumber \\ && +8 R_1 R_2 \frac{(R_1+R_2)^2}{L^4} + \frac{6 R_1 R_2}{L^2} \ \left( l\cdot \left( n_2 \frac{2 R_2}{L} - n_1 \frac{2 R_1}{L} \right)\right)^2       + {\cal O}((R/L)^5)\ . 
\eea

The two cross ratios have the same leading term $4R_2R_2/L^2$. The leading term of the difference is 
\be\label{eq:leadingT}
\chi_2-\chi_1\sim \frac{16 R_1^2 R_2^2}{L^4}\, (2 (n_1\cdot l)(n_2\cdot l)-n_1\cdot n_2)\,.  
\ee
The combination $(2 (n_1\cdot l)(n_2\cdot l)-n_1\cdot n_2)$ is always positive because $\chi_2\ge \chi_1$.   In fact, one can check that for two space-like separated causal diamonds,
\be\label{eq:condT}
(2 (n_1\cdot l)(n_2\cdot l)-n_1\cdot n_2) \ge 1\,,
\ee
by e.g. putting the four points in the same plane.

We assume the mutual information falls as a power of the distance $L$ and the leading term has a definite tensorial structure. This follows from the OPE in terms of field operators which we review in the next section. A power of $\chi_1$ or $\chi_2$ will give us the leading term power falloff, while the possible tensorial dependence can only come from an integer power of the difference $\chi_2-\chi_1$.
Therefore, while the overall power of $L$ of the leading contribution can be any, the tensorial character can only be an integer power of $(2 (n_1\cdot l)(n_2\cdot l)-n_1\cdot n_2)$, greatly simplifying the analysis when the contribution comes from interchange of fields with spin. That is, the long distance leading term of the mutual information has to be of the form
\be
I\sim c_k   \, (2 (n_1 \cdot l) (n_2\cdot l)-n_1\cdot n_2)^k\,\left( \frac{R_1^{2} R_2^{2}}{L^{4}} \right)^\Delta= C(n_1,n_2,l) \left( \frac{R_1^{2} R_2^{2}}{L^{4}} \right)^\Delta \,, \label{co2}
\ee
for some integer $k\ge 0$, and $c_k > 0$. Typically, there is a combination of different $k$ for the same $\Delta$ in the leading long $L$ limit. In App.~\ref{app:blocks} we compare this expression with the conformal block expansion of the mutual information. In this expansion $2\Delta$ and $k$ have the interpretation of the conformal dimension and spin of the field interchanged by the two spheres. In the rest of this section we investigate the restrictions on $\Delta$ for each $k$ 
 using mutual information strong superadditivity.

\subsection{Strong superadditivity for two spheres}\label{subsec:ssa2}

\begin{figure}[ht]
\begin{center}
\includegraphics[scale=0.2]{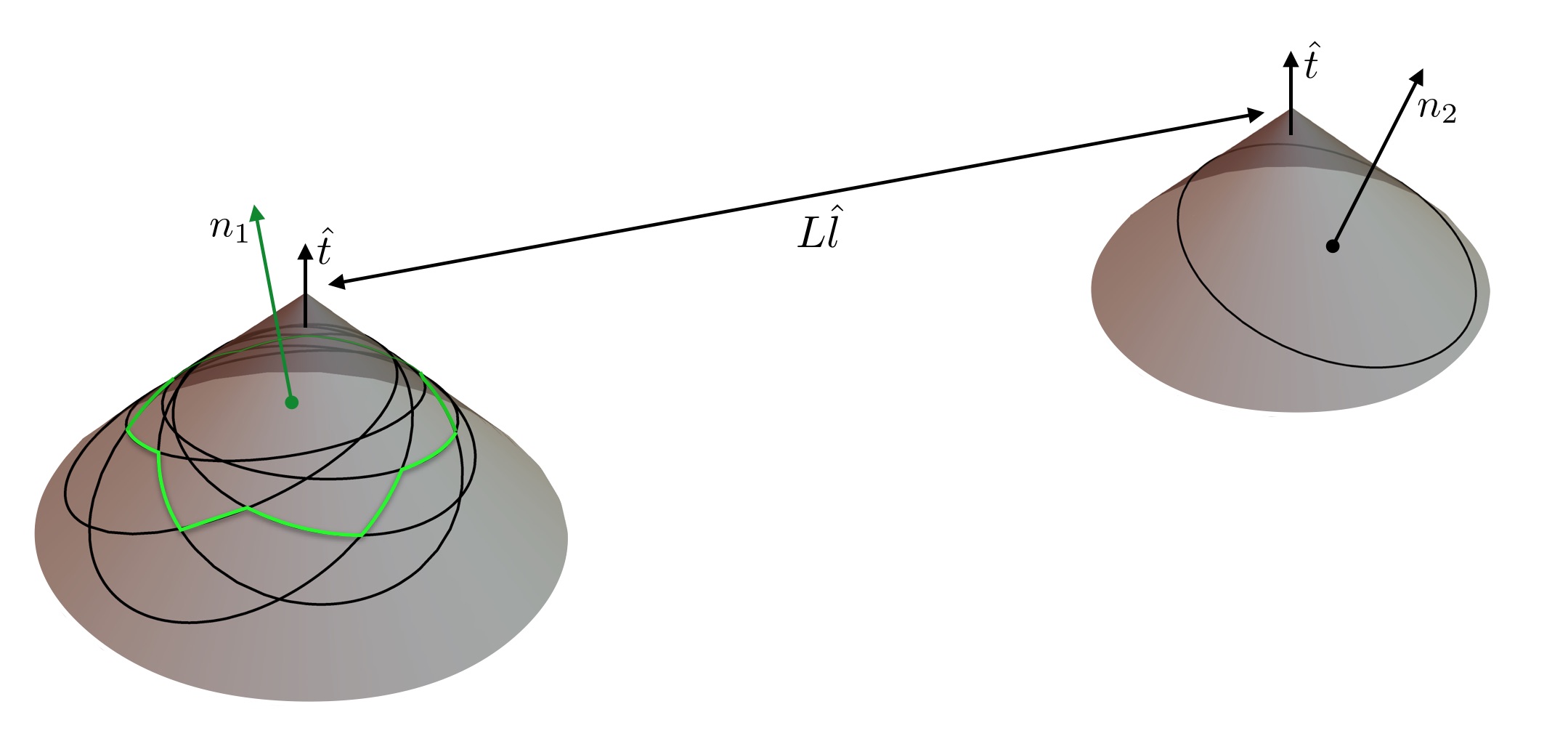}
\caption{Geometric setup of multiple boosted and rotated spheres used to derive (\ref{I}).} \label{setupspheres}
\end{center}
\end{figure}

We want to use strong superadditivity to get inequalities for the mutual information of two spheres. Therefore we need to have spheres on both sides of the inequality. This is possible only by taking many boosted rotated spheres and applying strong superadditivity many times. In the limit an inequality involving only spheres is obtained. The geometric construction is the same as the one used for proving the irreversibility theorems \cite{Casini:2012ei,casini2017markov,casini2018all}. See Fig.~\ref{setupspheres}.      

We will apply strong superadditivity changing the position of the first sphere in a null cone. The second sphere is kept fixed with radius $R_2$ and orientation $n_2$. The vector $x_3-x_1=L\, l$ is also fixed.\footnote{There are other inequalities for the mutual information coming from strong subadditivity for the entropy of the union of the two spheres when we change the position of the two spheres at the same time. However, these inequalities do not use the Markov property of the entropy of single spheres in the null cone and turn out to give weaker inequalities on $\Delta$. } 
 Following \cite{Casini:2012ei}, we take the radius of the first sphere   
$\tilde{r}_1=\sqrt{R_1 r_1}$ and rotate it around a fixed time-like unit vector $n_1$. We take many evenly rotated spheres.   
The rotated spheres have future directed time like unit vector
\be
\tilde{n}_1=\Lambda \frac{1}{2\sqrt{R_1 r_1}}(R_1+r_1,(R_1-r_1)\vec{\Omega})\,,\label{vector}
\ee
where  $\tilde{n}_1^2=-1$ (recall we are using the metric signature $(-1,1,\ldots,1)$), and $\vec{\Omega}$ is the unit spatial vector which later is averaged over the unit sphere. The average direction of the rotated spheres is given by
\be
n_1=\Lambda \hat{t}\,,\label{simple}
\ee
where $\hat{t}$ is the unit vector in the time direction and the matrix $\Lambda$ is a boost transformation that sets the average orientation of the spheres.      

Following \cite{Casini:2012ei} from the application of strong subadditivity we get
\be\label{eq:ISSA1}
 \frac{1}{\textrm{vol}(S^{d-2})}\int d\Omega \,  I(\t r_1, \t n_1)
 \le \int_{r_1}^{R_1} dr\,  \beta(r)\,  I(r, n_1)\,,
\ee
where we are only showing explicitly the dependence on the parameters of the left region of Fig.~\ref{setupspheres}.
The normalization factor $\textrm{vol}(S^{d-2})=2 \pi^{(d-1)/2}/\Gamma[(d-1)/2]$ is the volume of the unit sphere of spatial directions. The numerical coefficient appearing in the integral on the right hand side of (\ref{I}) is the density of the number of spheres as a function of the radius \cite{Casini:2012ei}: 
\be
\beta(r)=\frac{2^{d-3}\Gamma(\frac{d-1}{2})}{\sqrt{\pi}\Gamma(\frac{d-2}{2})}\frac{(r_1 R_1)^{\frac{d-2}{2}}(r-r_1)^{\frac{d-4}{2}}(R_1-r)^{\frac{d-4}{2}}}{r^{d-2}(R_1-r_1)^{d-3}}\,. 
\ee
 
For strong superadditivity (as happens with strong subadditivity) there is no loss of generality in considering only the infinitesimal version,\footnote{Strong subadditivity is equivalent to monotonicity of mutual information, and it is clear that monotonicity under inclusion is equivalent to its infinitesimal version.}  provided that all infinitesimal deformations of arbitrary regions are considered. Here we limit ourselves to the infinitesimal version of (\ref{eq:ISSA1}).  We can take $R_1=r_1+\epsilon$, and expand to quadratic order in $\epsilon$. In this approximation
\be
\tilde{n}_1 =\Lambda \left(\left(1+\frac{\epsilon^2}{8 r_1^2}\right)\hat{t}+\left(\frac{\epsilon}{2 r_1}-\frac{\epsilon^2}{4 r_1^2}\right) \hat{\Omega}\right)\,,
\ee
We will also use
\be
\int d\Omega\, \Omega^{i} = 0\,,\hspace{1cm}\int  d\Omega\, \Omega^{i} \Omega^{j} =   \frac{\textrm{vol}(S^{d-2})}{d-1}\,  \delta_{ij}\,.
\ee

Let us evaluate both sides of (\ref{eq:ISSA1}) to order $\epsilon^2$. The integrand in the left hand side expands to
\bea
I(\t r_1, \t n_1) & \approx& I(r_1, n_1) + \frac{\epsilon}{2} \partial_{r_1} I(r_1, n_1) + \epsilon^2 \Big[ \frac{1}{8 r_1} \left(r_1 \partial_{r_1}^2  I(r_1, n_1)- \partial_{r_1} I(r_1, n_1) \right)\\
& & + \frac{1}{8 r_1^2}n_1^\mu \partial_{n_1^\mu}I(r_1, n_1)+\frac{1}{4r_1} (\Lambda \Omega)^\mu \partial_{r_1}\partial_{n_1^\mu}I(r_1, n_1)+ \frac{1}{8 r_1^2}(\Lambda \Omega)^\mu (\Lambda \Omega)^\nu \partial_{n_1^\mu}\partial_{n_1^\nu}I(r_1, n_1)\Big] \nonumber\,.
\eea
Performing the angular integral gives\footnote{This amounts to replacing $(\Lambda \Omega)^\mu (\Lambda \Omega)^\nu \to \frac{1}{d-1}(g^{\mu\nu}+ n_1^\mu n_1^\nu)$.}
\bea
 \int \frac{d\Omega}{\textrm{vol}(S^{d-2})} \, I(\t r_1, \t n_1) & \approx& I(r_1, n_1) + \frac{\epsilon}{2} \partial_{r_1} I(r_1, n_1) + \epsilon^2 \Big[ \frac{1}{8 r_1} \left(r_1 \partial_{r_1}^2  I(r_1, n_1)- \partial_{r_1} I(r_1, n_1) \right) \nonumber\\
& +&  \frac{1}{8 r_1^2}n_1^\mu \partial_{n_1^\mu}I(r_1, n_1)+ \frac{1}{8 r_1^2}\frac{1}{d-1}(g^{\mu\nu}+ n_1^\mu n_1^\nu)
\partial_{n_1^\mu}\partial_{n_1^\nu}I(r_1, n_1)\Big]\,.\,
\eea
To evaluate the right hand side, we expand $I(r, n_1)$ to quadratic order in $r-r_1$, but keep the full dependence in $\beta(r)$ (to deal properly with the limit $r \to r_1$ and $r \to R_1$). The result is
\be
\int_{r_1}^{R_1} dr\,  \beta(r)\,  I(r, n_1) \approx I(r_1, n_1) + \frac{\epsilon}{2} \partial_{r_1} I(r_1, n_1)+ \frac{\epsilon^2}{8(d-1)r_1} \left(d r_1 \partial_{r_1}^2 I(r_1, n_1)-2(d-2) \partial_{r_1} I(r_1, n_1) \right)\,.
\ee
Therefore the mutual information strong superadditivity (SSA) for multiple spheres gives
\be\label{eq:SSA-general}
r_1^2 \partial_{r_1}^2I(r_1, n_1) -(d-3) r_1\partial_{r_1}I(r_1, n_1)- (g^{\mu\nu}+ n_1^\mu n_1^\nu)
\partial_{n_1^\mu}\partial_{n_1^\nu}I(r_1, n_1)-(d-1) n_1^\mu \partial_{n_1^\mu}I(r_1, n_1) \ge 0\,.
\ee

Before turning to the long distance limit, let us analyze the geometrical character of (\ref{eq:SSA-general}).
Using the covariant derivative
\be
\nabla_\mu = \frac{\partial}{\partial n^\mu} + n_\mu n^\nu\frac{\partial}{\partial n^\nu}
\ee
(this takes into account that $(n^\mu)^2=-1$ and so that $n^\mu \nabla_\mu=0$), we see that the angular variations combine to give the Laplacian,
\be
g^{\mu\nu} \nabla_\mu \nabla_\nu = (g^{\mu\nu} + n^\mu n^\nu) \frac{\partial}{\partial n^\mu}   \frac{\partial}{\partial n^\nu} +(d-1) n^\mu  \frac{\partial}{\partial n^\mu} \,.
\ee
Furthermore, we can characterize the causal diamond for a region in terms of an oriented area,
\be
A^\mu = r_1^{d-2} n_{1}^\mu\,\;,\; (n_1^\mu)^2=-1\,.
\ee
Then (\ref{eq:SSA-general}) becomes simply
\be
g^{\mu\nu} \frac{d}{dA^\mu}\frac{d}{dA^\nu} I(\chi_1, \chi_2) \ge 0\,.
\ee
For comparison, the infinitesimal strong subadditive inequality for the EE in $d$ dimensions is~\cite{casini2017markov}
\be
r^2 \Delta S''(r) - (d-3) r \Delta S'(r) \le 0\,,
\ee
which can be written as the second derivative with respect to the norm of the oriented area $|A| = r^{d-2}$. Compared to the EE, the mutual information inequality has angular dependence because we have two spheres.

\subsubsection{The long distance limit}

Let us now evaluate (\ref{eq:SSA-general}) in the long distance limit (\ref{co2}). We get
\be
2 (2\Delta+k)\left( \Delta-\frac{d+k-2}{2}\right) \left(2 (n_1 \cdot l) (n_2\cdot l)-n_1\cdot n_2\right)^k +k(k-1)\left(2 (n_1 \cdot l) (n_2\cdot l)-n_1\cdot n_2\right)^{k-2}\ge 0\,.\label{espre}
\ee
As a cross-check of the steps leading to (\ref{eq:SSA-general}), we can also compute (\ref{eq:ISSA1}) using the limit (\ref{co2}) from the start,
\be
 \frac{(r_1 R_1)^\Delta  R_2^{2\Delta}}{(\textrm{vol}(S^{d-2}))}\int d\Omega \,  C(\tilde{n}_1,n_2, l)
 \le  C(n_1,n_2,l)\,\int_{r_1}^{R_1} dr\,  \beta(r)\, r^{2\Delta}\, R_2^{2\Delta}\,. \label{I}
\ee
This leads to the same result (\ref{espre}).

Choosing spheres boosted with respect to $l$ with a large boost parameter $\sim\beta$ we get 
\begin{equation}
( 2 (n_1 \cdot l)(n_2 \cdot l)- n_1\cdot n_2 ) \sim \cosh(\beta).
\end{equation}
Therefore the first term in (\ref{espre}) has to be positive by itself. This gives
\be
\Delta\ge \frac{d+k-2}{2}\,.\label{torre}
\ee
This implies in particular that there is a lower bound on the decay of the mutual information at large distances for any CFT
\be
I\gtrsim L^{-2(d-2)}. 
\ee
We have obtained these bounds for spheres, but by monotonicity the same bounds hold for regions of any other shapes in the long distance limit. 
We will see in the next section how these bounds reproduce unitarity bounds for field operators.

In the long distance limit, the angular dependence is only through the combination
\be
T \equiv  2 (n_1 \cdot l) (n_2\cdot l)-n_1\cdot n_2\,.
\ee
Then (\ref{eq:SSA-general}) simplifies to
\be\label{eq:mSSAlong}
r_1^2 \partial_{r_1}^2I(r_1, T) -(d-3) r_1\partial_{r_1}I(r_1, T)- (T^2-1) \partial_T^2 I(r_1, T)-(d-1) T \partial_T I(r_1, T) \ge 0\,.
\ee
For (\ref{co2}), this readily reproduces (\ref{espre}). We find that (\ref{eq:mSSAlong}) can be expressed entirely in terms of cross-ratios in the long distance limit:
\be
u^2 \partial_u^2 I(u,v)+u \partial_v^2 I(u,v)+u(v-1) \partial_u \partial_v I(u,v)+\frac{4-d}{2} u\partial_u I(u,v)+\frac{2-d}{2}(v-1)\partial_v I(u,v) \ge 0\,.
\ee
But away from the long distance limit, we get that the inequality cannot be written solely in terms of cross-ratios.

\subsubsection{The case $d=2$}

Let us consider the mutual  SSA (strong superadditivity)
in the simplest case of $d=2$, varying the endpoints of interval $A$ as in the entropic C-theorem \cite{Casini:2004bw, Casini:2012ei}. Expanding to quadratic order in $\epsilon$ and using the chain rule, we obtain the differential inequality
\be\label{eq:SSAd2}
u^2 \partial_u^2 I + u v \partial_v^2 I+ \left(u^2+u(v-1) \right) \partial_u \partial_vI + u ( \partial_u I + \partial_v I) \ge0\,.
\ee
We see that the inequality can be expressed entirely in terms of the cross-ratios (\ref{eq:uv0}); this is special to $d=2$.

As a check, let us consider the long distance limit, where
\be
u \approx 16 \frac{R_1^2 R_2^2}{L^4}\;,\;v \approx 1- 8 \frac{R_1 R_2}{L^2} (2 (n_1 \cdot l) (n_2 \cdot l) - n_1 \cdot n_2)\,.
\ee
Replacing 
\be
I \approx \left(\frac{R_1^2 R_2^2}{L^4} \right)^\Delta \left( 2 (n_1 \cdot l) (n_2 \cdot l) - n_1 \cdot n_2\right)^k = u^\Delta\, \left(\frac{1-v}{2u^{1/2}} \right)^k
\ee
into (\ref{eq:SSAd2}) gives
\be
\left(\Delta^2- \frac{k^2}{4} \right) \left( 2 (n_1 \cdot l) (n_2 \cdot l) - n_1 \cdot n_2\right)^k+ \frac{1}{4} k(k-1) \left( 2 (n_1 \cdot l) (n_2 \cdot l) - n_1 \cdot n_2\right)^{k-2} \ge 0\,.
\ee
This is the right result for $d=2$.

\subsection{Saturation of inequalities, pinching property, and free fields}\label{subsec:saturate}

\begin{figure}[ht]
\begin{center}
\includegraphics[scale=0.3]{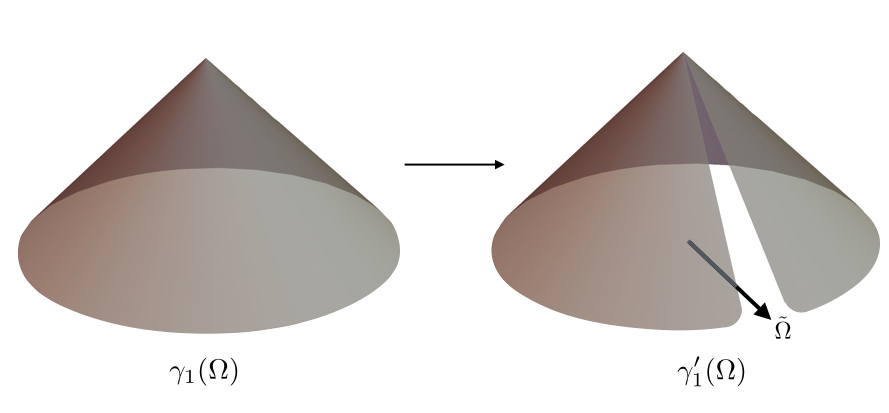}
\caption{Pinching $\gamma_1$ to $\gamma_1'$ gives a null surface with vanishing space-time volume.} \label{deform}
\end{center}
\end{figure}

The saturation of the inequality (\ref{torre}) gives a scaling dimension that coincides with the dimension of conformal primary free fields of helicity $h=k/2$.\footnote{See e.g.~\cite{Minwalla:1997ka} for a discussion of unitarity bounds and further references. We will return to this in Subsec. \ref{subsec:h}.} There is a surprising way in which saturation of strong superadditivity points to free fields in this context. To show this, we consider the following limit of the mutual information. As before, we take one of the regions to have boundary described by $\gamma_1(\Omega)$ on a null cone, where $\Omega$ are the angle variables describing the directions on the cone, and $\gamma_1(\Omega)$ is the radial (or temporal) coordinate on the surface. Consider deforming $\gamma_1(\Omega)$ only for a small region around some null direction $\tilde{\Omega}$, and taking the limit $\gamma_1(\tilde{\Omega})\rightarrow 0$ to a new surface $\gamma_1'(\Omega)$; see Fig.~\ref{deform}. In the limit when the region we cut out includes the apex of the causal cone, the space-time volume determined by the causal region associated to $\gamma_1'(\Omega)$ (the volume of the causal development of the corresponding null surface) is zero. The full causal region in this limit is just a null surface. We refer to this geometric deformation as ``pinching'' the original spacetime region.

It is known that smearing fields on a null surface is not enough to produce an operator in the Hilbert space, unless the operator is free \cite{schlieder1972some,Wall:2011hj,Bousso:2014uxa}. Therefore, if the theory does not contain free fields, the algebra associated to the region disappears in the limit where the first region approaches $\gamma_1'$, and the mutual information associated to $\gamma_1'$ should vanish. We call this property the ``pinching property'', that is
\be
\lim_{\gamma_1\rightarrow \gamma_1'}I(\gamma_1,B)=0\,, \hspace{1cm}\textrm{(pinching property: non-free models)}
\ee
for any fixed $B$. 
 In contrast, for free fields, the limit should give a non vanishing mutual information since there is a non trivial algebra for null surfaces in this case. Moreover, in the limit where the angular region which was pinched is very small there should be no change in the mutual information with respect to the original region, even if the space-time volume of the region vanishes. This is because, for a free field,  any operator in the region with boundary $\gamma_1$ can be expressed as an operator on the null surface by the equations of motion. Then the mutual information should be continuous as we take limits of regions on the null surface.     

Therefore, there is a very different behaviour of the mutual information as a function of $\gamma_1$ for free and non free theories. 
The continuity of the entropy, or non pinching property, gives a definition of what is a free model from the point of view of the entanglement entropy. Notice that this is not
exactly the same as the definition of free theory in terms of Wick's theorem for correlators. This latter applies for example to generalized free scalar fields with dimension $\Delta>(d-2)/2$ and in this case we expect the pinching property to hold.  
 
We will show that when $k=0, \Delta=(d-2)/2$ or $k=1, \Delta=(d-1)/2$ the mutual information is continuous under pinching on the null cone and the theory has to be free. For higher values of $k\ge 2$ again the saturation is associated to free fields but we discuss these cases later. For $k\ge 2$ and $\Delta=(d+k-2)/2$ the inequality (\ref{espre}) actually does not saturate and other terms with smaller $k$ have to appear with the same scaling $\Delta$ for the saturation of strong superadditivity to take place.    

Let us then consider $k=0, \Delta=(d-2)/2$ or $k=1, \Delta=(d-1)/2$. Inserting these values in the inequality (\ref{I}) (the non-infinitesimal inequality) it is the result of a direct calculation that it saturates exactly. As the inequality is the result of applying strong superadditivity many times for boosted and rotated spheres, it follows that the saturation of the final inequality implies the saturation of each of the intermediate instances where strong superadditivity was applied. In particular, strong superadditivity for the mutual information with just two boosted spheres must saturate. These spheres can be of any size and boost parameter. The following reasoning shows that this saturation implies in turn the saturation of strong superadditivity for any two surfaces $\gamma_a$, $\gamma_b$, on the light cone. We are only considering the long distance leading term.    
 
Let us consider the mutual information $I(A,B)$ with $B$ fixed and different $A$'s on a null cone. Strong superadditivity is
\bea
F(A_1,A_2)&=&I(A_1\cap A_2,B)+ I(A_1\cup A_2,B)- I(A_1,B)-I(A_2,B)\nonumber \\
&=&S(A_1 B)+S(A_2 B)-S((A_1\cap A_2)B )-S((A_1\cup A_2)B )\ge 0\,. \label{hy}
\eea
If we keep the intersection $A_1\cap A_2$ fixed, this quantity is decreasing with decreasing $A_1$ or $A_2$ by strong subadditivity. Therefore, if $F(A_1,A_2)=0$, the same holds for smaller regions with the same intersection. In the same way (\ref{hy}) is decreasing with increasing $A_1$ or $A_2$ keeping the union $A_1\cup A_2$ fixed.  Therefore, if $F(A_1,A_2)=0$, the same holds for bigger regions with the same union. Applying these two rules we get that $F(\gamma_a,\gamma_b)=0$  for any intermediate regions $\gamma_a,\gamma_b$ such that $A_1\cap A_2 \subset \gamma_a,\gamma_b \subset A_1\cup A_2$, and $\gamma_b\cap A_1 \subset \gamma_a\cap A_1$, $\gamma_a\cap A_2 \subset \gamma_b\cap A_2$. Given that we can choose spheres $A_1$, $A_2$ (with the same radius) arbitrarily, it is not difficult to see that we have $F(\gamma_a,\gamma_b)=0$ for any regions in the light cone.    

 Therefore, in analogy with the case of saturation of strong subadditivity of the entropy, we can say that in this situation the mutual information $I(\gamma,B)$ (in the long distance limit) is Markovian with respect to the surface $\gamma$ in the light cone 
 \be
 I(\gamma_1\cap \gamma_2,B)+ I(\gamma_1\cup \gamma_2,B)- I(\gamma_1,B)-I(\gamma_2,B)=0\,.
\ee 
This implies $I(\gamma,B)$ is a local functional of $\gamma(\Omega)$ \cite{casini2018all}, i.e. an integral     
  \be
  I(\gamma,B)=\int d\Omega\,  f(\gamma(\Omega),B)\,.
\ee  
 This directly implies the mutual information is non-pinching in the above sense. The model then has to contain a free field.    
We will see the cases $k=0$ or $k=1$ precisely correspond to free scalars and fermions respectively, and we will discuss in more detail the interpretation of $k>1$.

Away from the long distance limit the mutual information for free fields is of course non-pinching, but the inequality does not saturate any more. This is because of the contribution of polynomials of the free fields, as explained in the next sections. 

\section{Replica computation of the long distance expansion}
\label{replica}

In this section we derive a general formula for the mutual information coefficients in the long distance limit. The formula generalizes previous ones in the literature that were valid for spheres to the case of arbitrary regions. It involves the modular flows of the regions $A,B$. 

\subsection{OPE expansion of the mutual information}

We follow \cite{Cardy.esferaslejanas} to write the coefficient of the mutual information for well separated regions. The Renyi entropies of integer index are given by expectation values of twist operators in an $n$ replicated theory \cite{calabrese2004entanglement}. That is, 
\be
\tr \rho_A^n=\langle \Sigma_A^{(n)}\rangle\,,\label{rhs}
\ee 
where some regularization is assumed, which is equivalent to a small smearing of the operator $\Sigma_A^{(n)}$. This relation was originally thought in Euclidean time but also holds in real time for arbitrary space-like regions \cite{casini2012positivity}. The operator $\Sigma_A^{(n)}$ is just the twist (the operator that implements a symmetry locally) of the replicated theory corresponding to the symmetry of cyclic permutation of copies. This symmetry is evidently unbroken in the replicated model. The twist operator transforms  fields in the causal development of $A$ and commutes with the operators in the causal complement $A'$.    
The expectation value of the right hand side of (\ref{rhs}) is on the product state of the vacuum in each of the $n$ copies (or any other state formed by products of the same state in the different copies). For two disjoint regions $A$, $B$, we have 
\be
\tr \rho_{AB}^n=\langle \Sigma_A^{(n)} \Sigma_B^{(n)}\rangle\,.
\ee
Then, the idea of \cite{Cardy.esferaslejanas} is to replace each twist operator in this expression by an operator product expansion (OPE) in terms of field operators in each copy to obtain a series expansion of the mutual information for long distances. The OPE can be written 
\be
\Sigma^{(n)}_A=\tr \rho_A^n \, \sum_{\{k_j\}} C^{A\, \alpha_0\ldots \alpha_{n-1}}_{\{k_j\}}\, \prod_{j=0}^{n-1} \phi_{k_j}^{\alpha_j}(r_A^j)\,. \label{coco}
\ee
The pre-factor $\tr \rho^n_A$ is a normalization such that the first coefficient of the expansion corresponding to the identity operator is one, $C_1^{A}=1$. Each of the operators $\phi^{\alpha_j}_{k_j}$ belongs to the $j^{th}$ copy, $k_j$ determines the type of field operator and $\alpha_j$ is a possible spin index. $r_A^j$ is a point inside $A$ in the copy $j$. 

The Renyi entropy is $S_n(A)=(1-n)^{-1}\textrm{tr}(\rho_A^n)$, and the  Renyi mutual information
\be
I_n(A, B)=S_n(A)+S_n(B)-S_n(AB)=\frac{1}{1-n} \log \left(\frac{\tr \rho_A^n \, \tr \rho_B^n}{\tr \rho_{AB}^n} \right)\,.
\ee
In the long distance limit this is dominated by the operator of smallest dimension in the OPE expansion. The mutual information is obtained in the limit $n\rightarrow 1$. The leading term of the mutual information  is dominated by two copies of the smallest dimension operator in the theory, each one in a different replica.\footnote{The contribution to the mutual information of only one operator in one copy vanishes because the coefficient of the one copy operators in the expansion should have a factor $n-1$, such that $\lim_{n\rightarrow 1} \Sigma_A^{(n)}=1$, see \cite{agon2016quantum}.}   One then gets for the leading term
\bea
I(A,B) &\sim & \lim_{n\rightarrow 1}\left( \frac{n}{2 (n-1)} \sum_{j=1}^{n-1} C_{0j}^{A, \alpha \alpha'} C_{0j}^{B, \beta\beta'}\right) G_{\alpha\beta}(r_A, r_B) G_{\alpha'\beta'}(r_A, r_B)\label{yyy}\\
&=& D_{A,B}^{\alpha\alpha',\,\beta,\beta'}  G_{\alpha\beta}(r_A, r_B) G_{\alpha'\beta'}(r_A, r_B)\,,\nonumber
\eea
where 
\be
 G_{\alpha\beta}(r_A, r_B)=\langle 0| \phi^{\alpha}(r_A)\,\phi^{\beta}(r_B)  |0 \rangle\,.
\ee
If the operators in the two copies are complex there is an additional factor $2$ in (\ref{yyy}). 

Calling $\Delta$ to the dimension of the lowest dimension operator, (\ref{yyy}) falls as $L^{-4 \Delta}$. 
The coefficients could in principle be obtained from (\ref{coco}) taking correlators with fields at large space-like distances
\be
 C^{A \alpha \alpha'}_{0j}G_{\alpha\gamma}(r_A, r) G_{\alpha'\gamma'}(r_A, r) = \frac{\langle \Sigma_A^{(n)} \phi_0^\gamma(r) \phi_j^{\gamma'}(r) \rangle}{\tr \rho_A^n }\,, \hspace{1cm} r^2\rightarrow \infty\,.\label{ifi}
\ee

\subsection{Computation of the leading coefficient in terms of modular flows}

In order to better understand this coefficient we recall that (in presence of a regulator) the Hilbert space can be decomposed as ${\cal H}_A\otimes {\cal H}_{A'}$ for the region $A$, where $A'$ is the complementary region. Then we can write a tensor product basis for the Hilbert space on each copy as $\{|e^i_a,f^i_b\rangle\} $, where $i$ is the index of copies and $a,b$ the ones spanning the basis for each copy. Then we can write the twist operator in explicit form \cite{casini2012positivity}
\be
\Sigma_A^{(n)}=\bigotimes_{i=0}^{n-1} \left(\sum_a |e^{i+1}_a\rangle \langle e^i_a |\right )\otimes 1_{{\cal H}_{A'}}\,,     
\ee
with cyclic notation $|e^{n}_a\rangle\equiv |e^0_a\rangle$. This operator is unitary and does not depend on the chosen basis. If $|0^i\rangle$ is the global state in the copy $i$, and $|\Omega \rangle=\otimes_i |0^i\rangle $, it is not difficult to show using the Schmidt decomposition for $|0^i\rangle$ in ${\cal H}_A^i\otimes {\cal H}_{A'}^i$, that
\bea
\langle \Omega |\Sigma_A^{(n)} |\Omega\rangle &=& \tr \rho_A^n\,,\nonumber\\
\langle \Omega |\Sigma_A^{(n)} \,{\cal O}^{A,j}|\Omega\rangle &=& \tr ( \rho_A^n  \,{\cal O}^A)\,,\\
\langle \Omega |\Sigma_A^{(n)}  \,{\cal O}_1^{A,j} \, {\cal O}_2^{A,j'}|\Omega\rangle &=& \tr (  \,{\cal O}_1^{A}\, \rho_A^{n-(j-j')} \,  {\cal O}_2^{A}\,\rho_A^{(j-j')} )\,,\nonumber
\eea
where ${\cal O}^{A}$ is an operator in $A$ in the original single copy space,  and ${\cal O}^{A,j}$ is its representative in the $j^{th}$ copy. In the right hand side of these formulas we have expressions in a single copy Hilbert space. We also have 
\be
\Sigma_{A'}^{(n)}\Sigma_A^{(n)} |\Omega\rangle=|\Omega\rangle\,,\hspace{1cm}  \Sigma_A^{(n)} |\Omega\rangle=(\Sigma_{A'}^{(n)})^{-1}|\Omega \rangle\,,
\ee
where $(\Sigma_{A'}^{(n)})^{-1}$ just changes the order of copies with respect to $\Sigma_{A'}^{(n)}$.

Therefore, the right hand side of (\ref{ifi}), where the operators live in $A'$, amounts to evaluate
\be
\frac{\tr (\rho_{A'}^n \,\rho_{A'}^{-j} \phi^\gamma(r)\, \rho_{A'}^{j} \, \phi^{\gamma'}(r)  )}{\tr \rho_A^n }=\frac{\tr (\rho_{A'}^n \, \phi_A^\gamma(r,\tau=i j) \, \phi^{\gamma'}(r)  )}{\tr \rho_A^n }
\ee
where all operators are in a single copy theory, and 
\be
\phi_A^{\gamma}(r,\tau)=\rho_{A'}^{i \tau}  \phi^{\gamma}(r) \rho_{A'}^{-i \tau} 
\ee
is the field transformed by the modular flow of $A'$, which coincides, except for the sign of $\tau$, with the modular flow of $A$ (and this is why we have called it with subscript $A$).

Next we follow similar steps as in \cite{agon2016quantum}. It is convenient to write the sum over $j$ in (\ref{yyy}) as a contour integral in the complex plane. We write 
\bea
&&\sum_{j=1}^{n-1}\tr (\rho_{A'}^n \, \phi^\gamma_A(r_1,i j) \, \phi^{\gamma'}(r_1)  ) \,\,\tr( \rho_{B'}^n  \, \phi_B^\delta(r_2,i j) \, \phi^{\delta'}(r_2))\nonumber \\
&&\qquad \qquad=\oint d\tau \,\,\tr (\rho_{A'}^n \, \phi^\gamma_A(r_1,\tau) \, \phi^{\gamma'}(r_1)  ) \,\,\tr( \rho_{B'}^n  \, \phi_B^\delta(r_2,\tau) \, \phi^{\delta'}(r_2)) F(n,\tau)\,.\label{315}
\eea
The function $F$ is
\bea
F(n,\tau)=\frac{1}{2\pi i}\sum_{j=1}^{n-1} \frac{1}{\tau-i j}=\frac{1}{2\pi}(\psi(n+i \tau)-\psi(1+i \tau))\,,
\eea
where $\psi(x)=\Gamma'(x)/\Gamma(x)$ is the digamma  function. The contour of integration has to contain the imaginary points $i j$, $j=1,\ldots, n-1$ and avoid the origin. For $n>1$ this function has poles on the positive imaginary axis $\tau=i u$, $u\ge 1$. Then the contour of integration encircles this semi-axis to get an analytic continuation in $n$ of the sum (\ref{315}). In the limit $n\rightarrow 1$ we have
\be
F(n,\tau)
\rightarrow (1-n) \left(\frac{1}{2\pi}\psi'(-i \tau)+\frac{\pi}{2 \sinh^2(\pi \tau)} \right)+{\cal O}((n-1)^2)\,.
\ee
The first function within the brackets does not have poles along the positive imaginary axis and does not contribute. For the second term the integration contour  can be deformed to have $\tau=i/2  + s$, with $s$ real.   Then 
\bea
&&\lim_{n\rightarrow 1} (1-n)^{-1}\sum_{j=1}^{n-1}\tr (\rho_{A'}^n \, \phi^\gamma_A(r_1,i j) \, \phi^{\gamma'}(r_1)  ) \,\,\tr( \rho_{B'}^n  \, \phi_B^\delta(r_2,i j) \, \phi^{\delta'}(r_2))\nonumber \\
&&\qquad=-\int_{-\infty}^\infty ds\,\frac{\pi}{2 \cosh^2(\pi \,s)}\langle \phi^\gamma_A(r_1,i/2+s)\,  \phi^{\gamma'}(r_1)  )  \rangle\,\langle \phi^\delta_B(r_2,i/2+s)\, \phi^{\delta'}(r_2)  )  \rangle\,.
\eea

Putting all together, from (\ref{yyy}) and (\ref{ifi}), the coefficient of the correlators in the expansion of the mutual information in (\ref{yyy}) can be extracted from 
\bea
&& D_{A,B}^{\alpha\alpha',\,\beta,\beta'}  G_{\alpha\gamma}(r_A, r_1) G_{\alpha'\gamma'}(r_A, r_1) G_{\beta\delta}(r_B, r_2) G_{\beta'\delta'}(r_B, r_2)  \label{319}\\
&& \qquad =\int_{-\infty}^\infty ds\,\frac{\pi}{4 \cosh^2(\pi s)}\langle \phi^A_\gamma(r_1,i/2+s)\,  \phi_{\gamma'}(r_1)  )  \rangle\,\langle \phi^B_\delta(r_2,i/2+s)\, \phi_{\delta'}(r_2)  )  \rangle\,,\hspace{1cm} r_{1,2}^2\rightarrow \infty\,.\nonumber
\eea
Though we have used arguments about twist operators and reduced density matrices in the intermediate steps of the calculation, which require regularization, the final result (\ref{319}) makes perfect sense in the continuum theory itself. The correlation function of a field with a modular evolved field is analytic in the strip $\textrm{Im} \,\tau \in (0, 1)$, and obeys KMS periodicity between the boundaries of this strip \cite{Haag:1992hx}. Therefore there is no  divergence of the correlators for $s=0$ due to a suppression factor produced by the imaginary component of the modular parameter.
 
 This formula gives the coefficient for any shape of $A,B$.  However, it depends on the correlators with modular evolved fields, by the modular flows corresponding to $A,B$, in the far away limit. These modular flows are in general not easy to compute, except for the case of spheres, or other regions in the null cone, if we restrict attention  to the action of the flow on the null cone.  For spheres our formula coincides with the one in \cite{agon2016quantum}.   

\section{Evaluation of mutual information coefficients for spheres}
\label{examples}

In this section we compute the leading term of the mutual information directly from (\ref{319}). We recover some results in the literature for spheres as a check of the formulas in the previous section. We generalize these results to include the dependence of the mutual information with the relative boost between the spheres, as well as give the general result for arbitrary tensor structure of the primary fields. After analysing a few simple examples, we will focus on the contribution from general conformal primary fields. In particular, for spatial spheres the result is that of a scalar operator of the same conformal dimension times the dimension of the Lorentz representation.

We will need  formulas for the modular flow of spheres. Taking a sphere of radius $R$ centred at the origin with time-like direction $n=\hat{t}$, the modular flow is a conformal transformation that writes  (see \cite{Haag:1992hx}) 
\bea
x^0(\tau) &=&  N(\tau)^{-1} R \left(x^0 R \cosh(2 \pi \tau)+\frac{1}{2} (R^2-x^2) \sinh(2 \pi \tau)\right) \nonumber   \,,\\
x^i(\tau) &=& N(\tau)^{-1} \, R^2\, x^i     \,,\\
N(\tau) &=&  x^0 R \sinh(2 \pi \tau)+\frac{1}{2} \cosh(2 \pi \tau) (R^2- x^2)+\frac{1}{2}(R^2+ x^2) \nonumber\,,
\eea
with $x^2=-(x^0)^2+x^i x^i$. In formula (\ref{319}) we will need to evaluate this transformation for a far away spatial point $x^2\gg R^2$, that is otherwise arbitrary. For simplicity, will take $x^0=0$, or $x\cdot n=0$. We will also need to replace a complex modular parameter $\tau\rightarrow i/2+s$. After this change of parameter, and in the large $r=|\vec{x}|$ limit, we get simply 
\be
x(i/2+s)\rightarrow 0\,.
\ee  
This might be surprising, but is explained by the fact that the imaginary modular time $i/2$ effects an inversion of coordinates and sends the distant point $x$ to the origin. 
The coordinate transformation matrix is, to leading order in $r/R$, 
\be
\frac{dx^{\mu}(i/2+s)}{dx^\nu}= \Omega\, (2\, n^\mu n_\nu + I^\mu_ \nu(x))\,,  \label{mat}
\ee
where the conformal factor is
\be
\Omega= \frac{R^2}{r^2\, \cosh^{2}(\pi s)}\,,\label{omega1}
\ee
and 
\be\label{eq:Idef}
I^\mu_\nu(x)=\delta^\mu_\nu- 2 \hat{x}^\mu \hat{x}_\nu
\ee
is the matrix that implements inversions.\footnote{In other words, for ${x'}^\mu=x^\mu/x^2$, we have $\frac{\partial {x'}^\mu}{\partial x^\nu}=\frac{1}{x^2} I^\mu_\nu(x)$.}  
The Lorentz transformation in (\ref{mat})
\be
\Lambda^\mu_\nu=2\, n^\mu n_\nu + I^\mu_ \nu(x)\label{lambda}
\ee
changes the sign of the time coordinate and the spatial coordinate parallel to $\hat{x}$. It satisfies $\Lambda^2=1$. It corresponds to a boost in the direction $\hat{x}$ with imaginary boost  parameter $i \pi$.

\subsection{Scalar field}

We use formula (\ref{319}) for a real scalar field and spheres.  We have 
\be
 D_{A,B}= \lim_{r_{1,2}^2\rightarrow \infty} |r_1|^{4 \Delta}  |r_2|^{4 \Delta}  
 \int_{-\infty}^\infty ds\,\frac{\pi}{4 \cosh^2(\pi s)}\langle \phi_A(r_1,i/2+s)\,  \phi(r_1)  )  \rangle\,\langle \phi_B(r_2,i/2+s)\, \phi(r_2)  )  \rangle \,,\label{333}
\ee
where we have normalized the two point function of the scalar as $\langle \phi(0)\phi(x)\rangle=|x|^{-2 \Delta}$. In this formula the two expectation values are independent and we can use an independent coordinate system for each sphere where the spheres are centered at the origin, the orientation vectors are $\hat{t}$ and choose $r_{1,2}=r$, $r \cdot \hat{t}=0$.  

The modular evolved correlators write
\bea
  \langle \phi(r,\tau) \phi(r)\rangle &=& \Omega^{\Delta}(d^2(\tau))^{-\Delta}\,,\\
  d^2(\tau) &=& (x(\tau)-x)^2\,.
\eea
After the replacement $t\rightarrow i/2+s$ and the limit of $r\gg R$ we have
\be
d^2(\tau)\rightarrow r^2\,.\label{d}
\ee
Plugging this back in (\ref{333}), and using (\ref{omega1}), we get
\be
D_{A,B}= R_A^{2 \Delta} \, R_B^{2 \Delta}\, \int_{-\infty}^{\infty} ds\, \frac{\pi}{4}\frac{1}{\cosh(\pi s)^{2+4 \Delta}}= c(\Delta)\,R_A^{2 \Delta} \, R_B^{2 \Delta}\,.
\ee
with the numerical coefficient
\be
c(\Delta)=\frac{\sqrt{\pi} \, \Gamma[1+2\Delta]}{4\,\Gamma[3/2+2\Delta]}\,.\label{cs}
\ee
The mutual information becomes
\be
I(A,B)\sim c(\Delta)\frac{R_A^{2 \Delta} \, R_B^{2 \Delta}}{L^{4\Delta}}\,.\label{scal}
\ee
This result coincides with the one in \cite{agon2016quantum}; see also \cite{Calabrese:2010he} for intervals and \cite{Herzog:2014fra} for a numerical analysis of a specific case.

\subsection{Spinor field}

For a spinor field
\be
\langle \psi(0) \bar{\psi}(x)\rangle=i\frac{\hat{x}\!\!\!/}{|x|^{2\Delta}}\,.
\ee
The transformation law of the field is
\be
\psi = \Omega^\Delta\, S\left(\frac{1}{\Omega}\frac{\partial x'^\mu}{\partial x^\nu}\right) \psi'\,,
\ee
where $S(\Lambda)$ is the spinorial representation of the Lorentz transformation.  
 For $(\ref{lambda})$ this is the boost with imaginary argument $i \pi$ in the $\hat{x}$ direction:
\be
S(\Lambda)=-i\, (\not \!  n)(\not \!  \hat{x})\,.
\ee
Using this in (\ref{319}) we get
\be
D_{AB}= 2 c(\Delta) R_A^{2 \Delta}R_B^{2 \Delta}\, \not \!  n_1 \cdot \not \! n_2\,.\label{yyz}
\ee
The mutual information follows by contracting with two correlators:
\be
I(A,B)\sim   2^{[\frac{d}{2}]+1} c(\Delta) \frac{R_A^{2 \Delta} \, R_B^{2 \Delta}}{L^{4\Delta}} (2(n_1\cdot l)(n_2\cdot l)-(n_1\cdot n_2))\,.
\label{fer}\ee
For spheres in the same plane it gives the scalar result times $2\times 2^{[\frac{d}{2}]}$. The same result (without attention to the tensorial structure for relatively boosted spheres) was obtained in \cite{chen2017mutual}.

The result (\ref{yyz}) may strike for its simplicity, considering that in each region we can form different fermion bilinears with the same dimension
\be
\bar \psi \psi\;,\; \bar \psi \gamma^\mu \psi\;,\; \bar \psi [ \gamma^\mu, \gamma^\nu] \psi\;,\; \ldots
\ee
However, the first term does not contribute for a massless field,\footnote{In even dimensions, the contribution from $\bar \psi \psi$ vanishes due to chiral symmetry. In odd dimensions a nonzero contribution would violate parity and so is absent on the sphere.} and the antisymmetric tensors cannot contribute because they must be contracted with powers of the same vector $n^\mu$, or metric tensors. Therefore, only the vector contribution from $\bar \psi \gamma^\mu \psi$ survives. 
This agrees with a calculation in~\cite{Chen:2016mya}.

\subsection{Vector field}

Let's consider a primary field of spin 1,
\be
\la \phi_\mu(x) \phi_\nu(y) \ra = \frac{I_{\mu\nu}(x-y)}{|x-y|^{2\Delta}}\,,
\ee
where $I_{\mu\nu}$ was defined in (\ref{eq:Idef}). Recall that if a 
conformal transformation is decomposed into a local dilatation times a local Lorentz rotation,
\be
\frac{\partial x'^\mu}{\partial x^\nu} = \Omega(x)\, \Lambda^\mu_\nu(x)\,,
\ee
where
\be
\Omega^2= \frac{1}{d}\eta^{\alpha \nu} \eta_{\mu \beta} \frac{\partial x'^\mu}{\partial x^\alpha}\frac{\partial x'^\beta}{\partial x^\nu}\,,
\ee
then a primary spin one field transforms as
\be\label{eq:phiconf}
\phi_\nu = \Omega^\Delta\, \left( \frac{1}{\Omega} \frac{\partial x'^\mu}{\partial x^\nu}\right) \phi'_\mu\,.
\ee

In our present case, the conformal transformation comes from the modular flow, and the above gives
\be
\la \phi_\mu(r, \tau) \phi_\nu(r) \ra = \Omega^\Delta \left( \frac{1}{\Omega} \frac{\partial x'^\alpha}{\partial x^\mu}\right)\,\frac{I_{\alpha \nu}(x-x_0)}{(d^2(\tau))^{\Delta}}\,.
\ee
From (\ref{omega1}), (\ref{d}), and (\ref{mat}), we have to replace 
\bea
\la \phi_\mu(r, \tau) \phi_\nu(r) \ra &\rightarrow&  \frac{R^{2 \Delta}}{r^{4 \Delta} \cosh(\pi s)^{2 \Delta}} (2 n_\mu n^\alpha + I_\mu^\alpha(r)) I_{\alpha\nu}(r)\nonumber\\
&=&\frac{R^{2 \Delta}}{r^{4 \Delta} \cosh(\pi s)^{2 \Delta}} I_{\mu\delta}(r)(2 n^\delta n^\alpha + g^{\delta\alpha})I_{\alpha\nu}(r)\,.
\eea
In the last step we have used the orthogonality $n\cdot r=0$. 

Inserting this in (\ref{319}) the factors of $I_{\mu\nu}(r)$ cancel on both sides of the equation and  we get
\be
D_{AB}^{\mu_1\mu_2,\,\nu_1\nu_2} =D_{AB}^{\textrm{scalar}}  (2 n_1^{\mu_1} n_1^{\mu_2} + g^{\mu_1\mu_2}) (2 n_2^{\nu_1} n_2^{\nu_2} + g^{\nu_1\nu_2}) \,. \label{asa}
\ee
The expansion of the mutual information follows by contracting this tensor with two correlators with the indices $\mu_1,\nu_1$ and $\mu_2,\nu_2$. We get
\be
I(A,B)\sim  c(\Delta)\frac{R_A^{2 \Delta} \, R_B^{2 \Delta}}{r^{4\Delta}} \left[4\left(2(n_1\cdot l)(n_2\cdot l)-n_1\cdot n_2\right)^2 + (d-4)\right]  \,.\label{vector1}
\ee
Note that for spatial spheres, this simplifies to $d$ times the scalar contribution, which is the right number of components for a spin one field.

\subsection{Antisymmetric rank 2 field}\label{subsec:Fmunu}

We now analyze the case of a primary antisymmetric tensor field $F_{\mu\nu}$ of dimension $\Delta$ in $d$ space-time dimensions. A special example of interest is the Maxwell field with $\Delta=2$ in $d=4$, which is free. The following combinations at $A$ and $B$ contribute to the mutual OPE:
\be\label{eq:exchangeF}
F_{\mu\nu}F^{\mu\nu}\,, \;\; n^\mu n^\alpha\, F_{\mu\nu} F^\nu_\alpha \,.
\ee
This gives a maximum tensorial index $k=2$, but contains $k=0$ too. When inserted in the mutual SSA, the $k=2$ contribution gives rise to both $k=2$ and $k=0$ terms, see (\ref{espre}). The first one saturates when the field becomes free, but the second one does not. Since a free field should saturate exactly the mutual SSA, it has to be that the additional $k=0$ exchange in (\ref{eq:exchangeF}) does the job of cancelling the nonzero term in (\ref{espre}). Using the modular flow result (\ref{319}) we now calculate the contribution of this primary field to the mutual information. We use this result in the next section to exhibit the precise cancellation on the inequality when $d=4$ and $\Delta=2$. 

The correlator is constructed from the inversion tensor (\ref{eq:Idef}),
\be\label{eq:FFcorr}
\la F_{\mu \nu}(x) F_{\alpha \beta}(0) \ra = \frac{I_{\mu \alpha}( \hat x) I_{\nu \beta}( \hat x) - I_{\mu \beta}( \hat x) I_{\nu \alpha}( \hat x)}{|x|^{2\Delta}}\,,
\ee
with $\hat x^\mu =x^\mu/|x|$. Similarly to (\ref{319}), the conformal transformation from the modular flow in (\ref{319}) acts on $F_{\mu\nu}$ as a local dilatation and a local Lorentz transformation on each of the indices of the field.  For $r \gg 1$ as needed in  (\ref{319}), we have
\be\label{eq:Fboosted}
\la F_{\mu\nu}(r, \tau) F_{\alpha \beta}(r) \ra \to \frac{R^{2\Delta}}{r^{4\Delta} \cosh^{2\Delta}(\pi s)} \Lambda_\mu^{\mu'}(n) \Lambda_\nu^{\nu'}(n) \left(I_{\mu' \alpha}(r) I_{\nu' \beta}(r)-I_{\mu' \beta}(r) I_{\nu' \alpha}(r) \right)
\ee
where the Lorentz transformation is given in (\ref{lambda}).

Replacing (\ref{eq:Fboosted}) into (\ref{319}) gives
\bea
D_{A,B}^{\mu\nu,\mu'\nu';\,\rho \sigma, \rho'\sigma'} &=& \frac{1}{16} D_{A,B}^{\text{scalar}} \left(I_{\mu\mu'}(i n_1) I_{\nu\nu'}(i n_1)-I_{\mu\nu'}(i n_1) I_{\nu\mu'}(i n_1) \right) \nonumber\\
&&\qquad \times \left(I_{\rho \rho'}(i n_2) I_{\sigma \sigma'}(in_2)-I_{\rho \sigma'}(i n_2) I_{\sigma \rho'}(in_2) \right)\,.
\eea
Note that the inversion tensor is evaluated on the imaginary vector $i n_\mu$, $I_{\mu \nu}(i n)=2 n_\mu n_\nu + g_{\mu\nu}$.
Contracting with
\be
\la F_{\mu\nu}(r_A) F_{\rho\sigma} (r_B) \ra\,\la F_{\mu'\nu'}(r_A) F_{\rho'\sigma'} (r_B) \ra\,,
\ee
after a few manipulations we obtain the long distance limit of the mutual information
\be
I(A,B)\sim  c(\Delta)\frac{R_A^{2 \Delta} \, R_B^{2 \Delta}}{L^{4\Delta}} \left[4(d-2)\left(2(n_1\cdot l)(n_2\cdot l)-n_1\cdot n_2\right)^2 + \frac{(d-4)^2-d}{2}\right]  \,.\label{eq:IFmunu}
\ee
For spatial spheres, this gives $d(d-1)/2$ times the scalar contribution, again reproducing correctly the number of components in $F_{\mu\nu}$. 

\subsection{Contribution of an arbitrary conformal primary}\label{subsec:arbitrary}

After going over the previous examples, we will now analyze the contribution of a conformal primary field in a general tensor	 Lorentz representation. We will first obtain a formula for the case of boosted spheres in terms of the character of a particular Lorentz transformation. Specializing to spatial spheres we will find the remarkably simple result that a general conformal primary contributes as a scalar of the same dimension times the dimension of the spin representation. Using the formula for characters of the Lorentz group we then give the general result in an explicit form.

Consider a conformal primary field $\mathcal O_\alpha$ of dimension $\Delta$ transforming in a representation ${\cal R}$ of the Lorentz group. Here $\alpha$ is a multi-index for the vector space of the representation ${\cal R}$, and the representation corresponds to a Young tableaux determining the symmetry of the indices\footnote{For some brief information about Young diagrams see section \ref{subsec:h} below.}.  The propagator
\be\label{eq:OO}
\langle \mathcal O_\alpha(x) \mathcal O_\beta(0) \ra = G_{\alpha \beta}(x) = \frac{1}{|x|^{2\Delta}} \hat G_{\alpha \beta}(x)\,,
\ee
where $ \hat G_{\alpha \beta}(x)$, which contains the tensor structure, is independent of $\Delta$. See
\cite{Alkalaev:2012rg, Costa:2014rya} for correlation functions for mixed-symmetry tensors. In more detail, we can construct the two-point function in terms of products of inversion tensors 
\be
\mathcal I_{\alpha \beta}(x)\equiv I_{\alpha_1 \beta_1}(x)\ldots  I_{\alpha_n \beta_n}(x)\label{forma}
\ee
 projected onto the appropriate symmetry space,
\be
\hat G_{\alpha \beta}(x) = (P^{({\cal R})})_{\alpha}^{\;\bar \alpha} \,\mathcal I_{\bar \alpha \bar \beta}(x) \,(P^{({\cal R})})_{\;\beta}^{\bar{\beta}}=(P^{({\cal R})})_{\alpha}^{\;\bar \alpha} \,\mathcal I_{\bar \alpha  \beta}(x)=\mathcal I_{ \alpha \bar \beta}(x) \,(P^{({\cal R})})_{\;\beta}^{\bar{\beta}}\,,
\ee
where $P^{({\cal R})}$ is the corresponding projector. In the last two equations we have used that the projectors over definite Young-tableaux symmetry commute with the product of two index tensors such as (\ref{forma}).\footnote{In fact all operators of index permutation and traces commute with this type of tensors. In particular Lorentz transformations of the tensor space have the same structure (\ref{forma}) and commute with permutations and traces, keeping invariant the symmetry type of a tensor. }  This together with the property ${\cal I}^2=1$ gives
\be 
\hat{G}\cdot \hat{G}= P^{({\cal R})}\,,\label{cuadrado}
\ee
where we have used a matrix notation. 

As an example, for an antisymmetric tensor $F_{\mu\nu}$,
\begin{align}
\la F_{\mu\nu} F_{\rho \sigma} \ra &=\frac{1}{|x|^{2\Delta}} \,(I_{\mu\rho}(x)I_{\nu\sigma}(x)-I_{\mu\sigma}(x)I_{\nu\rho}(x)) \\
&= \frac{1}{|x|^{2\Delta}} P_{\mu\nu}^{\;\bar \mu \bar \nu} \,I_{\bar \mu \bar \rho}(x)I_{\bar \nu \bar\sigma}(x)\,P^{\bar \rho\bar \sigma}_{\;\rho \sigma}\,,
\end{align}
where the projector for the rank two antisymmetric representation is
\be
P_{\mu\nu}^{\;\bar \mu \bar \nu} =\frac{1}{2} \left( \delta_\mu^{\bar \mu} \delta_\nu^{\bar \nu}- \delta_\mu^{\bar \nu} \delta_\nu^{\bar \mu}\right)\,.
\ee

Given these formulas, let us analyze the leading contribution of a conformal primary operator to the mutual information,
\be
I(A, B) \sim D_{A,B}^{\alpha \alpha',\,\beta,\beta'}  G_{\alpha\beta}(r_A, r_B) G_{\alpha'\beta'}(r_A, r_B)\,,
\ee
with the coefficient $D_{A,B}$ computed in (\ref{319}) using the modular flow of the two-point function. We record this here for convenience:
\bea
&& D_{A,B}^{\alpha\alpha',\,\beta\beta'}  G_{\alpha\gamma}(r_A, r_1) G_{\alpha'\gamma'}(r_A, r_1) G_{\beta\delta}(r_B, r_2) G_{\beta'\delta'}(r_B, r_2)  \label{3191}\\
&& \qquad =\int_{-\infty}^\infty ds\,\frac{\pi}{4 \cosh^2(\pi s)} \langle \phi^A_\gamma(r_1,i/2+s)\,  \phi_{\gamma'}(r_1)  )  \rangle\,\langle \phi^B_\delta(r_2,i/2+s)\, \phi_{\delta'}(r_2)  )  \rangle\,,\hspace{1cm} r_{1,2}^2\rightarrow \infty\,.\nonumber
\eea
The modular flow in the right hand side acts in terms of a conformal transformation,
\begin{align}
\langle \phi^A_\gamma(r_1,i/2+s)\,  \phi_{\gamma'}(r_1)  )  \rangle &= \frac{\Omega^\Delta}{|d(\tau)|^{2\Delta}}\, R(\Lambda_1)_\gamma^{\,\bar \gamma} \hat G_{\bar \gamma \gamma'}(r_1) \nonumber\\
& \to  \left( \frac{R_A^2}{r_1^4\, \cosh^{2}(\pi s)}\right)^\Delta\,\, R(\Lambda_1)_\gamma^{\,\bar \gamma} \hat G_{\bar \gamma \gamma'}(r_1)\,.
\end{align}
Here $\hat G$ is defined in (\ref{eq:OO}), $R(\Lambda_1)$ is the Lorentz transformation (\ref{lambda}) in the  representation $R$ of the primary operator, and in the last line we used the rescaling factor $\Omega$ found above in (\ref{omega1}) as well as the limit $r_1 \to \infty$ that gives (\ref{d}).

Note that $\hat G$ is given in terms of products of inversion tensors projected onto the appropriate symmetry subspace, and similarly $R(\Lambda_1)$ will contain tensor products of $\Lambda^\mu_{1\,\nu}=2\, n_1^\mu n_{1\,\nu} + I^\mu_ \nu(r_1)$ times projectors. 
Recalling $r_1 \cdot n_1=0$, gives
\be
\Lambda^\mu_{1\,\nu}\,I_{\nu \rho}(r_1) = I_{\mu \rho} (i n_1)\,,
\ee
independent of $r_1$. Therefore
\be\label{eq:useful}
R(\Lambda_1)_\gamma^{\,\bar \gamma} \hat G_{\bar \gamma \gamma'}(r_1)= \hat G_{\gamma \gamma'} (i n_1)\,.
\ee
We found examples of this property previously for the current and the antisymmetric field. Using this and (\ref{cuadrado}), the solution to (\ref{3191}) simplifies to
\be
D_{A,B}^{\alpha\alpha',\,\beta\beta'}  = c(\Delta) (R_A R_B)^{2\Delta}\, \hat G^{\alpha \alpha'}(i n_1)\, \hat G^{\beta \beta'}(i n_2)\,.
\ee
The result for the leading mutual information is then
\begin{align}\label{eq:mutual-general}
I(A, B) &\sim c(\Delta)\left(\frac{R_A R_B}{L^2}\right)^{2\Delta}\,\hat G^{\alpha \alpha'}(i n_1) \hat G_{\alpha'\beta'}(l) \hat G^{\beta' \beta}(i n_2) \hat G_{\beta \alpha}(l)  \nonumber\\
&=c(\Delta)\left(\frac{R_A R_B}{L^2}\right)^{2\Delta}\, \Tr \left(\hat G(i n_1) \hat G(l) \hat G(i n_2) \hat G(l) \right)\,.
\end{align}
This general expression is quite simple and compact: we find the scalar field result times a factor that comes from conjugating the modular-boosted operator with the two-point function $\hat G(l)$ that connects the two regions. The trace is taken over the representation ${\cal R}$ of the conformal primary, and this part of the contribution is independent of $\Delta$.

We can give a geometric interpretation to this result by rewriting  it as
\bea
I(A, B) &\sim & c(\Delta)\left(\frac{R_A R_B}{L^2}\right)^{2\Delta}\, \Tr \left(P^{({\cal R})}\cdot {\cal I}(i n_1)\cdot {\cal I}(l)\cdot {\cal I}(i n_2)\cdot{\cal I}(l)\right)\label{rpp} \\& = & c(\Delta)\left(\frac{R_A R_B}{L^2}\right)^{2\Delta}\, \Tr_{{\cal R}} \left( \Lambda(n_1,l,n_2,l) \right)\,. \nonumber
\eea
Since ${\cal I}(i n_a)$, $a=1,2$, is the reflection in the plane perpendicular to $n_a$, and ${\cal I}(l)$ is the one in the plane perpendicular to $l$, the matrix $\Lambda(n_1,l,n_2,l)$ in the trace is a Lorentz transformation formed by the product of four reflections. Then, the mutual information is proportional to the character of the representation ${\cal R}$ evaluated for a particular, geometrically determined, Lorentz transformation.\footnote{The Lorentz transformation $\Lambda(n_1,l,n_2,l)$ is not invariant under cyclic permutation of the arguments in general, but the trace it is.}  

Before giving a more explicit expression, let us apply this result to spatial spheres.  In this case, $n_1= n_2$ and both are orthogonal to $l$, the unit vector in the direction of $r_A - r_B$. In this case the reflections in (\ref{rpp}) commute and cancel each other, giving $\Lambda(n_1,l,n_2,l)=1$. Then
 the trace is just the dimension of the representation, and we thus arrive to
\be
I(A, B)  \sim\,\text{dim}(\mathcal R)\, c(\Delta)\left(\frac{R_A R_B}{L^2}\right)^{2\Delta}\,.
\ee
We conclude that for spatial spheres the leading contribution of a conformal primary operator to the mutual information is that of a scalar field of the same conformal dimension times the dimension of its Lorentz representation. 

We can make more explicit the general expression (\ref{rpp}) using the characters of the representations of Lorentz group \cite{hirai1965characters}. These are given in terms of the eigenvalues of the Lorentz transformation. Calling   
\be 
\cosh(\beta)=2 (n_1 \cdot l)\, (n_2 \cdot l)-n_1 \cdot n_2\,,
\ee 
the eigenvalues of $\Lambda(n_1,l,n_2,l)$ (the product of four reflections) are  $(e^{2 \beta}, e^{-2 \beta} , 1 ,\ldots,1)$. A quick way to derive this is to evaluate $\Lambda(n_1,l,n_2,l)$ when the four points $x_i$ of Fig. \ref{two_spheres} are placed on the same plane. Here we present the result, while technical details of the derivation are in appendix \ref{cuentas}. 

For odd dimensions $d=2 q+3$, $q=0,1,\ldots$, the Young diagram giving the representation of the Lorentz group is determined by the lengths $m_1,\ldots,m_{q+1}$ of the rows, with $0\le m_1\le m_2\le \ldots m_{q+1}$.\footnote{If the number of rows in the Young diagram is less than $q+1$ we have to complete the sequence with zeros.} Defining the matrices 
\be
  A_j = 
\left\lbrace\frac{(m_s+s-1/2)^{2 p-1}}{(2 p-1)!}\right\rbrace_{p=1,\ldots,q}  
^{s=1,\ldots, \hat{j},\ldots , q+1 } \,,\hspace{.5cm} (\textrm{for}\, d=3,\, q=0,\, A_1=\{1\})\,,
\ee
where $\hat{j}$ is to be omitted,  we have
\be\label{eq:oddI}
I(A, B)  \sim\, c(\Delta)\left(\frac{R_A R_B}{L^2}\right)^{2\Delta}  \left(\frac{\sum_{j=1}^{q+1} (-1)^{j+q+1}\,\sinh(2 \beta (m_j+j-1/2))\, \det(A_j) }{\sinh(\beta) (\cosh(2 \beta)-1)^q}\right)\,. 
\ee

For even $d=2 q+2$, $q=1,2,\ldots$ we have instead
\bea
A_j & = &  
\left\lbrace\frac{(m_s+s-1)^{2 p}}{(2 p)!}\right\rbrace_{p=0,\ldots,q-1}  
^{s=1,\ldots, \hat{j},\ldots , q+1 } \,,\\
I(A, B)  &\sim&\, c(\Delta)\left(\frac{R_A R_B}{L^2}\right)^{2\Delta}  \left( \frac{  \,\sum_{j=1}^{q+1} (-1)^{j+q+1}\,\cosh(2 \beta (m_j+j-1))\, \det(A_j) }{(\cosh(2 \beta)-1)^q}\right)\,. \label{iff}
\eea
For even dimensions, if $m_1\ne 0$, we have two dual representations with the same tensor structure. If both components are present we have to multiply (\ref{iff}) by $2$.

Let us apply this result to some examples. In the case of a graviton in $d=4$ ($q=1$), the lowest dimension primary is the curvature tensor, having symmetry given by a square Young diagram with $m_1=2,m_2=2$ and $\Delta=3$. This gives
\be
I(A, B)  \sim\, \frac{512}{3003}\left(\frac{R_A R_B}{L^2}\right)^{6}  \left(32\, (2 (n_1 \cdot l)\, (n_2 \cdot l)-n_1 \cdot n_2)^4-24\,   (2 (n_1 \cdot l)\, (n_2 \cdot l)-n_1 \cdot n_2)^2 +2\right)\,. 
\ee  
Another example is the contribution for a symmetric two index tensor, corresponding to $\vec{m}=(0,\ldots,0,2)$. We get
\bea
I(A, B)  &\sim &  c(\Delta) \left(\frac{R_A R_B}{L^2}\right)^{2\Delta} \,\times \\
&& \hspace{-2cm} \left(16\, (2 (n_1 \cdot l)\, (n_2 \cdot l)-n_1 \cdot n_2)^4+4 (d-6)\,   (2 (n_1 \cdot l)\, (n_2 \cdot l)-n_1 \cdot n_2)^2 + \frac{d(d-7)}{2}+7\right)\,.\nonumber
\eea
The case of the stress tensor follows replacing $\Delta=d$. The case of a totally antisymmetric field of $d/2$ indices in even dimensions reads
\be
I(A, B)  \sim   c(\Delta) \left(\frac{R_A R_B}{L^2}\right)^{2\Delta} \, 4\left(\begin{array}{c} d-2\\ \frac{d-2}{2}\end{array}\right)\left( \, (2 (n_1 \cdot l)\, (n_2 \cdot l)-n_1 \cdot n_2)^2- \frac{1}{d}\right)\,.
\ee
For $\Delta=d/2$ this corresponds to the free helicity $1$ field. 

\section{Consequences of the inequalities and the structure of the mutual information}\label{sec:consec}

In section \ref{subsec:saturate} we argued that the strong superadditive inequality saturates, in the long distance limit and on the null cone,  precisely for free fields. 
The formula for the long distance expansion of the mutual information allows us to prove this in a different way.  
Then we discuss the interpretation of the inequalities in terms of unitarity bounds. The bounds provided by mutual information superadditivity are sharp only for fields with spin structure that supports a free conformal field.    The saturation of the inequality for free fields will allow us to give the general form of the long distance expansion for two arbitrary regions with boundaries $\gamma_A$ and $\gamma_B$ on null cones.

\subsection{Free fields have  leading coefficients which are local on the null cone}

For a region $A$ determined by a surface $\gamma_A$ on the null cone, the modular flow for a CFT acts independently and locally on each null line of the cone, while it is non local outside of it \cite{casini2017modular}. For a free field, using the linear equations of motion,  we can write the field $\phi^\gamma(r)$ as a linear expression on the same field smeared on the null cone.\footnote{We have to conformally extend Minkowski space to the cylinder, such that the cone extends to a complete Cauchy surface.} Then, the operator formed by its modular evolution $\phi^\gamma_A(r,i+s)$ depends on $\gamma_A$ as a sum over the different null rays. In consequence, the coefficient $D_{A,B}^{\alpha,\alpha',\beta,\beta'}$ of formula (\ref{319}) is additive over the different null rays that form the region $\gamma_A$. In particular the mutual information is Markovian with respect to $\gamma_A$ in the long distance limit, in the sense that strong subadditive combinations vanish. 

This exactly coincides with our findings in section \ref{subsec:saturate} on the saturation of the strong superadditivity inequality precisely for the dimensions of free fields.  For an interacting field, the field equations that allow us to write the field as an operator on the null cone are non linear, and the mutual information is not additive as a function of the null rays in $\gamma_A$. There is also the fact that, strictly speaking, we cannot write the interacting field sharply on the null cone. 

 For free fields, composite fields formed by products of fields can also be written on the null cone but as non linear polynomials in the free fields at different points. Therefore, the subleading contributions produced by these composite fields are non local on the angular variables of the null cone and do not saturate the strong superadditive inequalities. Still, these contributions are given by multiple integrals over the light ray directions. Hence, they will also be non pinching. This is as expected, since a quenching of any subleading term in the mutual information of a free field as we pinch the null surface would give a non continuous behaviour which is not allowed for a free field. The functional form of the coefficients as a function of $\gamma_A$ in the interacting case must then be more subtle. In the interacting case, the coefficients of the expansion of the mutual information could not be written as convergent expansions of multiple integrals along the angular variables since these contributions are continuous under pinching.      

\subsection{Unitarity bounds}\label{subsec:h}

As described above, the inequalities can only saturate for free fields, and will more generally give unitarity bounds related to the free field dimensions. Let us describe these inequalities for the specific examples in section \ref{examples}. We recall the inequality 
\be
2 (2\Delta+k)\left( \Delta-\frac{d+k-2}{2}\right) \left(2 (n_1 \cdot l) (n_2\cdot l)-n_1\cdot n_2\right)^k +k(k-1)\left(2 (n_1 \cdot l) (n_2\cdot l)-n_1\cdot n_2\right)^{k-2}\ge 0\,.\label{espre1}
\ee
corresponding to a mutual information leading term $I\sim (R_1^2 R_2^2/L^4)^{\Delta}\,(2 (n_1 \cdot l) (n_2\cdot l)-n_1\cdot n_2)^k$. 

For a scalar field the structure of the mutual information (\ref{scal}) corresponds to the case $k=0$, which according to (\ref{espre1}) gives $\Delta\ge (d-2)/2$. This coincides with the unitarity bound of a scalar field. Since the mutual information measures correlations coming from all fields of the theory, any scalar field violating this bound would also violate strong superadditivity. For a Dirac fermion field the formula (\ref{fer}) corresponds to the case $k=1$. As for the case $k=0$, eq. (\ref{espre1}) has only one term in this case. The inequality gives $\Delta\ge (d-1)/2$ corresponding to the unitarity bound for a Dirac field.   

The mutual information for the antisymmetric field $F_{\mu\nu}$ is given by (\ref{eq:IFmunu}). In this case we have a combination of terms with $k=2$ and $k=0$ with the same $\Delta$.  These two contributions arise from the operators $F_{\mu\nu}F^{\mu\nu}\,, \;\; n^\mu n^\alpha\, F_{\mu\nu} F^\nu_\alpha $ contributing to the OPE in each of the regions. Accordingly, the inequality in this case is the corresponding linear combination of the inequalities (\ref{espre1}) for $k=2,k=0$:
\be\label{max}
4 (d-2)(2 \Delta -d) (\Delta +1) \left(2(n_1\cdot l)(n_2\cdot l)-n_1\cdot n_2 \right)^2 +((d-9) d+16) \Delta  \left(\Delta -\frac{d-2}{2}\right)+4 (d-2) \ge 0\,.
\ee
This follows from plugging (\ref{eq:IFmunu}) into the mutual SSA (\ref{eq:mSSAlong}). We recall that $(2 (n_1 \cdot l) (n_2\cdot l)-n_1\cdot n_2)\ge 1$, and can have arbitrarily large values for relatively boosted spheres. Then to satisfy the inequality we have to have a positive leading $k=2$ term, giving $\Delta\ge d/2$. However, the unitarity bound for an antisymmetric conformal primary field is $\Delta \ge d-2$, which is stronger. This coincides with $\Delta \ge d/2$ only for $d=4$, where the saturation corresponds to the free Maxwell field. 
 We can check that the full inequality (\ref{max}) is always satisfied, given the unitarity bound. However, it is in general weaker than the unitarity bound for an antisymmetric field, except for $d=4$, where the saturation coincides with the field being free.

This allow us to understand what will happen more generally for other free fields. For free scalar and fermions the equations (\ref{espre1}) for $k=0,1$ contain only one term and saturate at the unitarity bound. Higher spin free fields give place to larger $k$ in (\ref{espre1}), and the saturation of the unitarity bound makes the first term vanish but not the second. However, for these higher spin free fields there are other contributions to the inequality corresponding to terms with the same $\Delta$ but lower $k$. These terms must cancel in the full inequality that must saturate completely for a free field. In fact (\ref{max}) vanish for $d=4, \Delta=2$.

For the vector field mutual information (\ref{vector1}) we have again a combination of the cases $k=2$ and $k=0$.  Replacing into the mutual SSA inequality gives
\be\label{eq:mSSAJ2}
2(2\Delta-d)(\Delta+1) (2 (n_1 \cdot l)(n_2 \cdot l)-n_1 \cdot n_2)^2+\frac{1}{2} \left[4-(d-4) \left(d-2-2\Delta \right)\Delta \right] \ge 0\,.
\ee
The first term can be increased by boosting, so it gives $\Delta \ge d/2$. This is the unitarity bound for free helicity one fields. The unitarity bound for the vector field is instead $\Delta\ge (d-1)$, saturating for conserved currents.  This is in agreement with our previous discussion, because a conserved current is in general not a free field. Related to this, the second term with $k=0$ does not vanish either. On the other hand, when $d=2$ both of these terms vanish. This saturation indicates the presence of a free field. We can understand this by recalling that in this dimensionality the current is dual to the derivative of a free field, $J^\mu= \epsilon^{\mu\nu} \partial_\nu \phi$, and the conservation equation $\partial_\mu J^\mu=0$ becomes the Klein-Gordon equation for the scalar.\footnote{Using (\ref{eq:condT}) one can check that (\ref{eq:mSSAJ2}) is positive definite for all $(d, \Delta)$ with $\Delta\ge d/2$.
}

\begin{figure}[h!]
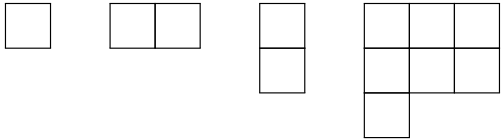

\begin{center}  
\ydiagram{1} \qquad \ydiagram{2}\qquad \ydiagram{1,1}\qquad \ydiagram{3,3,1}
\caption{Examples of Young diagrams corresponding to $J_\mu$, $T_{\mu\nu}$ (symmetric), $F_{\mu\nu}$ (antisymmetric) and $\Phi_{[\mu_1\mu_2\mu_3][\nu_1 \nu_2][\sigma_1 \sigma_2]}$. In this last case, we also symmetrize over $(\mu_1 \nu_1 \sigma_1)$ and over $(\mu_2 \nu_2 \sigma_2)$.}
\label{fig:young}
\end{center}  
\end{figure}

Let us now recall the situation in $d$ dimensions for general spin.\footnote{See e.g.~\cite{Mack:1975je, Minwalla:1997ka}. We also found very useful the more recent work \cite{Costa:2014rya}.} Conformal primary fields are classified by their dimension $\Delta$ and an irreducible representation of $SO(d)$. These can be encoded by Young diagrams, with a box for each tensor index, and boxes in a row (column) denoting symmetrization (resp. antisymmetrization) of the associated indices. The number of boxes in row $i$ is denoted by $l_i$, and the number of boxes in column $i$ is $h_i$. The diagram is ordered according to $l_1 \ge l_2 \ge \ldots \ge l_{h_1}$, and $h_1 \ge h_2 \ge \ldots \ge h_{l_1}$. The corresponding tensor, making manifest the antisymmetrizations, is
\be
\Phi_{[\mu^1_1 \ldots \mu^1_{h_1}][\mu^2_1 \ldots \mu^2_{h_2}] \ldots [\mu^{l_1}_1 \ldots \mu^{l_1}_{h_{l_1}}]}\,.
\ee
The tensor is traceless, and the overall height $h_1$ has to be smaller than the rank of the $SO(d)$ algebra ($d/2$ for $d$ even, or $(d-1)/2$ for $d$ odd).
We illustrate these properties in Fig. \ref{fig:young}.

\vskip 5mm

The unitarity bound is
\be\label{eq:ubound}
\Delta \ge l_1- h_{l_1}+d-1\,,
\ee
where, as just reviewed, $l_1$ is the length of the first (longest) row, and $h_{l_1}$ is the length of the rightmost (shortest) column. In particular, a completely symmetric tensor with $l$ indices obeys 
\be
\Delta \ge  d+ l -2\;\;,\;\; (l_1=l\,,\, h_1=1)\,,
\ee
while for a completely antisymmetric tensor with $h_1$ indices
\be
\Delta \ge d-h_1\;\;,\;\; (l_1=1\,,\, h_1)\,.
\ee
When the bound is saturated, a first level descendant of the primary operator becomes null; this translates into conservation equations $\partial_\mu \Phi^{\mu \ldots}=0$. 

Note that at fixed $l\equiv l_1$ the smallest unitarity bound in (\ref{eq:ubound}) corresponds to the largest possible $h_{l_1}$. This is attained when $h_1 = \ldots =h_l = \lfloor \frac{d}{2} \rfloor$, namely for a rectangular Young diagram with $ \lfloor \frac{d}{2} \rfloor$ rows and $l$ columns. This gives the smallest bound
\be\label{eq:smallest}
\Delta \ge d-1 -\left \lfloor \frac{d}{2} \right\rfloor + l\,.
\ee
For even space-time dimensions, the bound is saturated by free primary fields of helicity $l$, which have the tensor structure~\cite{Siegel:1988gd, Howe:1989vn, Minwalla:1997ka}
\be\label{eq:Phih}
\Phi_{[ \mu^1_1 \ldots \mu^1_{d/2} ] [ \mu^2_1 \ldots \mu^2_{d/2} ]\ldots [ \mu^l_1 \ldots \mu^l_{d/2} ]}\,.
\ee
Besides the conservation equation, free primary fields satisfy an extra constraint analogous to the Bianchi identity.\footnote{
A half-integer helicity corresponds to a spinor, and the number of blocks is then the floor of the helicity.} Besides free scalars and fermions, a familiar example is the Maxwell field $F_{\mu\nu}$ with $l=1$ and $d=4$, discussed above. In higher dimensions, the free helicity $l=1$ field is a $d/2$ form $F_{[\mu_1 \ldots \mu_{d/2}]}$. Another example is the Riemann curvature tensor $R_{\alpha\beta\gamma\delta}$ for a free graviton in $d=4$ (helicity $l=2$). For $d$ odd, free fields occur only for $l=0$ (Klein-Gordon scalar) and $l=1/2$ (free Dirac fermion).

We see that the saturation of the mutual superadditive inequality in $d$ dimensions for a given $k$ and $\Delta=(d-2+k)/2$ must correspond to a free field of helicity
\be
l = \frac{k}{2}\,.
\ee
(Note that $l$ here is what we have denoted by the helicity $h$ in other sections. We have not used this notation here to avoid confusion with the heights $h_i$ of the Young diagram.)
We have tensorial contributions with $k=2l,\,2l-2,\,\ldots 0$. The leading $k$ term has zero coefficient for the conformal dimension of the free field.   The other coefficients cancel each other such that the mutual superadditive inequality is exactly saturated. 

In this way the strong superadditive inequality gives the unitarity bound for all primary fields of the form  (\ref{eq:Phih}). It does not however give the most constraining bound on $\Delta$ for fields with other spin structure. The reason is that these fields saturate the unitarity bound without being free. On a more practical level, we see that the inequalities are labelled by just one number $k$, which is just enough to produce the bounds for the fields (\ref{eq:Phih}). Other fields with different index structure can only give place to the same type of leading $k$ contributions for the mutual information. The existence of free fields giving place to the same $k$ contribution is hiding the unitarity bound so that it is not visible from the superadditive inequality.

\subsection{General form of the coefficient on the null cone for free fields}\label{subsec:local-form}

The fact that the inequality must exactly saturate for a free field gives us the general structure of the coefficients in this case. For spheres, and helicity $h$, we must have an expansion of the form
\be
I(A,B)\sim c((d-2+2 h)/2) \,\textrm{dim}({\cal R}(h))\, \left(\frac{R_A R_B}{L^2}\right)^{(d-2+2 h)} \sum_{s=h_0}^{h} a_{s} \,(2 (n_1 \cdot l) (n_2\cdot l)-n_1\cdot n_2)^{2 s} \,,
\label{511} \ee
with $h_0=0$ or $h_0=1/2$ for integer or half-integer helicity respectively, and $\textrm{dim}({\cal R}(h))$ is the dimension of the representation of helicity $h$. This formula corresponds to a highest value of $k=2 h$ and $\Delta=(d-2+2 h)/2$. The saturation of (\ref{espre}) gives the relations between the coefficients
\be
a_{s}=-\frac{(d+2 h+ 2 s-4)(h-s+1)}{s(2 s-1)}\,a_{s-1}\,.  
\ee
For spatial spheres $(2 (n_1 \cdot l) (n_2\cdot l)-n_1\cdot n_2)=1$, and  we obtain from the general result for spatial spheres
\be  
\sum_{s=h_0}^h a_s= 1\,.
\ee
This completely fixes the $a_s$,
\be
a_s=\frac{b_s}{\sum_{l=h_0}^h  b_l}\,,\hspace{1cm}b_s=\prod_{l=h_0+1}^s -\frac{(d+2 h+ 2 l-4)(h-l+1)}{l(2 l-1)}\,,\hspace{.3cm}b_{h_0}=1.
\ee

Further, as we have argued before, the saturation of strong superadditivity tells us that the dependence of the mutual information on arbitrary regions $\gamma_A,\gamma_B$ with boundary on null cones must be additive with respect to the angular directions. This gives us the general form of the dependence in the shape of these regions. In fact, if we want to construct a tensor out of the surface described by $\gamma_A^\mu$, and that this tensor is formed additively on the different null lines, we can only form the quantities
 \be  
C_A^{\mu_1\mu_2\ldots \mu_n}=(\textrm{vol}(S^{d-2}))^{-1}\int d\sigma_x\, \gamma_A^{\mu_1}(x) \gamma_A^{\mu_2}(x)\ldots \gamma_A^{\mu_n}(x)\,,\label{cc}
 \ee
 where $\gamma_A^\mu(x)$ is the null vector position of the surface and $x$ parametrizes the null cone, and the normalization factor $\textrm{vol}(S^{d-2})$ is the volume of the $d-2$ dimensional unit sphere.
This tensor has dimension $n+d-2$, is completely symmetric and traceless. We need two such tensors for $A$ and $B$ and these tensors can only be contracted by two correlators, as shown by the replica trick calculation. Since the total dimension is zero we must have for a free field $n=2h$. 
 Since correlators of primary fields in a CFT are combinations of the inversion tensor $I_{\mu\nu}$, and these correlators are contracted with the symmetric tensors (\ref{cc}) we have the general structure for helicity $h$ in any dimensions
\be\label{eq:Ih}
I_h(A,B)\propto L^{-2(d-2+2h)}\, C_A^{\mu_1\mu_2\ldots \mu_{2h}} I_{\mu_1\nu_1}(l)I_{\mu_2\nu_2}(l)\ldots I_{\mu_{2h}\nu_{2h}}(l) C_B^{\nu_1\nu_2\ldots \nu_{2h}}\,,
\ee
where $l$ is the unit vector in the direction from $A$ to $B$. 

We can calibrate the overall coefficient of this expression in each case. For example, taking into account that for a sphere we have 
\be
C_A=R_A^{d-2}\,,\hspace{.5cm} C_A^\mu=R_A^{d-1} n_A^\mu\,,\hspace{.5cm}C_A^{\mu\nu}=R_A^{d}\, \frac{d}{d-1}\, (n_A^\mu n_A^\nu + g^{\mu\nu}/d)\,.
\ee 
and using (\ref{511}), we obtain for arbitrary regions with boundary in the null cone,
\bea
I(A,B)&\sim &  \frac{c(\frac{d-2}{2})}{L^{2(d-2)}}\, C_A C_B\,, \hspace{8.5cm}h=0\,, \\
I(A,B)&\sim &  \frac{c(\frac{d-1}{2})}{L^{2(d-1)}}\,2^{[\frac{d}{2}]+1}\, \left(2 (C_A^\mu l_\mu) (C_B^\nu l_\nu) - C_A^\mu C_{B\,\mu} \right)\,, \hspace{3.5cm} h=1/2\,,\\
I(A,B)&\sim &  \frac{c( \frac{d}{2})}{L^{2 d}}\,\frac{4(d-1)^2}{d^2}\left(\begin{array}{c}  {\small d-2} \\ \frac{d-2}{2} \end{array}\right) \left(4 (C_A^{\alpha\beta}l_\alpha l_\beta) (C_B^{\alpha\beta}l_\alpha l_\beta) -4 C_A^{\alpha\beta}C^B_{\alpha \sigma} l_\beta l^\sigma +C_A^{\alpha\beta}C^B_{\alpha \beta}\right)\,, \hspace{.0cm}h=1 \,.\nonumber \\
\eea

As discussed in App. \ref{app:blocks}, the mutual information for boosted spheres is a conformally-invariant function of the cross-ratios $(u,v)$, which can be expanded in terms of conformal blocks. These arise from the exchange of primary traceless symmetric operators between regions $A$ and $B$. Evaluating (\ref{eq:Ih}) for boosted spheres should then give the leading contribution from a conformal block of spin $2h$; the reason is that this is the contraction of $2h$ inversion tensors (which come from the two-point function of the exchanged primary) with the unique symmetric traceless tensor formed with $2h$ $n_i$'s on each sphere. Writing for this case
\be
C_A^{\mu_1 \ldots \mu_{2h}} = n_1^{\alpha_1} \ldots n_1^{\alpha_{2h}}\, P_{\alpha_1 \ldots \alpha_{2h}}^{\mu_1 \ldots \mu_{2h}}
\ee
where $P$ is the projector onto symmetric traceless rank $2h$ tensors (and similarly for region $B$), we obtain
\bea
I_h(A,B)&\propto& L^{-2(d-2+2h)}\,  n_1^{\alpha_1} \ldots n_1^{\alpha_{2h}}\, P_{\alpha_1 \ldots \alpha_{2h}}^{\mu_1 \ldots \mu_{2h}}I_{\mu_1\nu_1}(l)I_{\mu_2\nu_2}(l)\ldots I_{\mu_{2h}\nu_{2h}}(l) \, n_2^{\nu_1} \ldots n_2^{\nu_{2h}} \nonumber\\
&\sim & L^{-2(d-2+2h)}\, C_{2h}^{d/2-1}\left(n_1 \cdot I(l) \cdot n_2 \right)
\eea
where in the last step we have used the Gegenbauer polynomials $C_{2h}^{d/2-1}(x)$ (see eq. (2.17) of \cite{Costa:2011dw}). In App. \ref{app:blocks} we show that this agrees with the result from the conformal block expansion.

\section{Conclusions}\label{sec:concl}

The strong superadditive property of the mutual information for a CFT for regions with boundary on the null cone, Eq. (\ref{eq:mutualSSA}), is a consequence of strong subadditivity of the entropy plus the Markov property of the entropy on the null cone. This inequality implies certain unitarity bounds (\ref{torre}) when applied to the long distance expansion of the mutual information. These correspond to the unitarity bounds that saturate for free fields. These unitarity bounds, derived from mutual information properties al long distances, depend only on the two point function of the leading primary field. As a generalized free field with the relevant two point function of dimension $\Delta$ is a valid CFT, having the operator with dimension $\Delta$ as the lowest dimensional operator, the bounds apply to any field in any CFT, disregarding the presence of other lower dimensional fields.   

There is a natural reason why these are the only bounds captured by the mutual information inequality. Saturation of strong superadditivity implies the mutual information has a local expression as a function of the surface $\gamma$ on the null cone; see Subsec. \ref{subsec:local-form}. This in turn implies that the mutual information does not vanish under pinching of the surface along a null ray, i.e., a local deformation along a null ray that makes the spacetime volume enclosed in the causal region vanish (see Fig. \ref{deform}). The mutual information does not vanish under pinching only for free fields which have a non trivial algebra localizable on the null surface. This may be taken as an entropic definition of a free field. 

We have also shown a general formula (\ref{319}) for the long distance expansion of the mutual information for regions of arbitrary shapes that depends on the modular flows of these regions. This formula independently shows the locality of the long distance mutual information on the null cone for free fields. For spheres with arbitrary orientations we obtained in (\ref{rpp}), (\ref{eq:oddI}), (\ref{iff}), the general form of the leading long distance contribution from general primary  operators with spin. 

For spherical regions the contribution of specific primary fields to the  mutual information in the long distance limit is universal, in the sense that it depends only on the two point function of the field and not on the specific theory to which it belongs. This is because the modular flow for the spheres is geometrical and universal. Further, as the geometric form of the flow does not depend on the spacetime dimension, the coefficients do not depend on $d$ in an explicit form.   This is not the case for other regions. Hence it is expected that further details of the theory would appear into the mutual information coefficients in this more general case. We can see that this must be so for example considering the case of a conserved current, which has a universal two point function. The current can belong to a free theory, in which case the contribution has to be continuous under pinching, while the opposite must hold for the contribution of a current in an interacting theory.    

It is possible that the inequalities for non spherical regions and interacting models may contain additional useful information. However, it seems at present difficult to compute these coefficients in interacting theories. Another interesting question is how the information on the other unitarity bounds is encoded in the mutual information, if at all. A possibility would be that these bounds should result (in the entropy) from (unknown) inequalities beyond strong subadditivity involving higher derivatives with respect to the shape.

\section*{Acknowledgments} 
This work was partially supported by CONICET, CNEA, and Instituto Balseiro, Universidad Nacional de Cuyo, Argentina. H.C. acknowledges an ``It From Qubit" grant of the Simons Foundation. 
H.C. and G.T. are also supported by ANPCYT PICT grant 2018-2517. E. T. is supported by the Department of Energy grant DE-SC0019139. 

\appendix

\section{Conformal block expansion of the mutual information}\label{app:blocks}

In a CFT, it is useful to expand the mutual information in conformal partial waves,\footnote{Properties of conformal partial waves can be found for instance in~\cite{Dolan:2011dv} and references therein.}
\be\label{eq:cpw}
I(A,B) = \sum_{\lbrace \Delta, \ell \rbrace} b_{\Delta, \ell}\,G_{\Delta, \ell}(u,v)
\ee
where the cross-ratios are defined as in (\ref{eq:uv0}), which we reproduce here,
\be
u = \frac{x_{12}^2 x_{34}^2}{x_{13}^2 x_{24}^2}= \chi_1^2\;,\; v= \frac{x_{14}^2 x_{23}^2}{x_{13}^2 x_{24}^2}= \frac{\chi_1^2}{\chi_2^2}\,.
\ee
The conformal block contributions correspond to an exchange of a primary operator appearing in the expansion of the Renyi operators corresponding to the two spheres~\cite{long2016co, chen2017mutual, Chen:2016mya}. The contribution of each sphere must be proportional to the given primary field contracted with powers of the vectors $n_i$ that determine the orientation of the sphere. Contractions with the metric vanish because primaries are traceless. Therefore operators with some antisymmetric indices have zero coefficient in the expansion and only symmetric traceless representations appear in the conformal block expansion.  In this appendix we analyze the long distance limit of (\ref{eq:cpw}), obtaining the expansion used in (\ref{co2}). We will identify the parameters $\Delta, \ell$ in the conformal partial wave in terms of the leading operator contribution to the mutual information. At the end, we explore briefly the null limit for the mutual information.

The long distance limit for the mutual information (see Sec.~\ref{subsec:confinv}) corresponds to $u\to 0, v\to 1$, with the cross-ratios related to the geometric parameters by
\bea\label{eq:longuv}
u &=& 16 \frac{R_1^2 R_2^2}{L^4} + \mathcal O(L^{-5}) \nonumber\\
v &=& 1- 8 \frac{R_1 R_2}{L^2} ( 2 (n_1 \cdot l)\, (n_2 \cdot l)-n_1 \cdot n_2)+ \mathcal O(L^{-3})\nonumber\\
&=&1+2 u^{1/2} \hat x_{21} \cdot I(\hat x_{31}) \cdot \hat x_{43}\,.
\eea
In the last step we used
\be
\hat x_{21}^\mu I_{\mu\nu}(\hat x_{31}) \hat x_{43}^\nu= n_1^\mu (g_{\mu\nu}-2 l_\mu l_\nu) n_2^\nu=-( 2 (n_1\cdot l)\, (n_2 \cdot l)-n_1 \cdot n_2 )\,.
\ee
We see that the tensorial dependence from (\ref{eq:leadingT}) and (\ref{co2}) appears in this limit via $(1-v)/2u^{1/2}$. 

To relate the conformal partial wave expression to the long distance limit of Sec.~\ref{sec:ssa} we  need to take the simultaneous limit $u \to 0, \,v\to 1$. In this limit, the conformal partial wave is \cite{Costa:2011dw}
\be
\lim_{u\to0,v\to1}G_{\Delta, l}(u, v) \sim c_{d,l}\,u^{\frac{\Delta}{2}}\, C_l^{d/2-1}\left(\frac{v-1}{2u^{1/2}} \right)\,,
\ee
where $c_{d,l}$ is a normalization constant, and $C_l^{d/2-1}(x)$ is a Gegenbauer polynomial. For reference, the values for the first few spins are
\bea\label{eq:Cinstance}
C_0^{d/2-1}(x)&=&1\nonumber\\
C_1^{d/2-1}(x)&=&(d-2)x\nonumber\\
C_2^{d/2-1}(x)&=&\frac{d(d-2)}{2} (x^2-1/d)\nonumber\\
C_3^{d/2-1}(x)&=&\frac{d(d-2)}{6}((d+2)x^3-3x)\;.
\eea

Hence, we recognize that this reproduces the long distance limit with the tensorial dependence in (\ref{co2}), after identifying $\Delta$ here with $2 \Delta$ in (\ref{co2}), and $\ell$ with $k$.
The conformal partial wave expression also shows that the tensorial dependence includes the leading term in (\ref{co2}) as well as lower powers $(2 (n_1 \cdot l)\, (n_2 \cdot l)-n_1 \cdot n_2)^{k-2}, \ldots$, whose coefficients are uniquely fixed in terms of the leading one. This can also be checked explicitly. As an example, let us consider the exchange of a primary traceless symmetric $l=2$ operator. Its two-point function is
\be
\langle \phi_{\mu\nu}(x) \phi_{\rho \sigma}(0) \rangle = \frac{C_\phi}{x^{2\Delta}}\, \left (\frac{1}{2} I_{\mu\rho}(x) I _{\nu\sigma}(x)+\frac{1}{2} I_{\mu\sigma}(x) I _{\nu\rho}(x)- \frac{1}{d} g_{\mu\nu} g_{\rho \sigma} \right)
\ee
where the last term is fixed by the tracelessness condition. The contribution to the mutual information for two boosted spheres is obtained by contracting with the normal vectors $n_i$ of each region (contraction with the metric vanishes since the operator is traceless). This gives
\be
n_1^\mu n_1^\mu \,\langle \phi_{\mu\nu}(x) \phi_{\rho \sigma}(0) \rangle \, n_2^\rho n_2^\sigma \propto \left(2 (n_1\cdot l)\, (n_2 \cdot l)-n_1 \cdot n_2 \right)^2 - \frac{1}{d}\,,
\ee
which reproduces the relative coefficient predicted by the $l=2$ Gegenbauer polynomial in (\ref{eq:Cinstance}).

Finally, we ask if the null limit for the mutual information yields new bounds. This is obtained by boosting $x_3$ and $x_4$ relative to $x_1$ and $x_2$. This is accomplished by fixing $x_{14}^2$ (so that the two regions are space-like), while $x_{13}^2, x_{23}^2$ and $x_{24}^2$ become large. In this limit, $u \to 0$ and $v \to 0$, with $u/v^2$ fixed. When $u \to 0$ (cfr. eq. (2.39) in~\cite{Dolan:2011dv}),
\be\label{eq:Gsmallu}
G_{\Delta, \ell}(u,v) \sim u^{\frac{1}{2}(\Delta - \ell)}(1-v)^\ell\, {}_2F_1 \left( \frac{\Delta+\ell}{2}+a, \frac{\Delta+\ell}{2}+b; \Delta+ \ell; 1-v\right)\;,\;u \to 0\;,
\ee
where the parameters $a$ and $b$ are related to the dimensions of the conformal primary operators in the 4-point function, $a= - \frac{1}{2}(\Delta_1- \Delta_2)$, $b=\frac{1}{2}(\Delta_3- \Delta_4)$. In our case, we only need to consider $a=b=0$. Expanding the hypergeometric function in $G(u, v)$ for $v \to 0$  gives
\be
G_{\Delta, \ell}(u,v) \sim\,\frac{\Gamma(\Delta+\ell)}{\Gamma(\frac{\Delta+\ell}{2})^2}  u^{\frac{1}{2}(\Delta - \ell)}\,\log(1/ v)\,.
\ee
As before, we need to replace here $\Delta \to 2 \Delta,\, \ell \to k$ for the mutual information. Plugging this expression into (\ref{eq:SSA-general}) and taking the limit $n_1 \cdot l \gg 1$, $n_2 \cdot l \gg 1$ yields the unitarity bound $\Delta\ge \frac{d+k-2}{2}$. This is the same as the long distance expansion result, so in this respect the null limit is not adding new constraints. Nevertheless, it would be interesting in future work to perform a more detailed analysis of the mutual information in the null limit.

\section{Details on the calculation of the general coefficient for spheres}
\label{cuentas}
According to (\ref{rpp}) the calculation of the general coefficient of the mutual information for arbitrary spheres and representations boils down to the calculation of the
 character of the Lorentz representation for a specific Lorentz transformation. We use the formulas for the characters in \cite{hirai1965characters} to get an explicit result. These characters are functions of the eigenvalues of the Lorentz transformation matrix. For the Lorentz transformation in question (see section \ref{subsec:arbitrary}) the eigenvalues  are 
 \be
 1,\ldots,1,e^{2 \beta},e^{-2 \beta}\,\label{eie}.
 \ee
 For these particular values the formulas for the characters have an indeterminate $0/0$ expression, and we have to take a limit.  
 The formulas for the characters are different according to the parity of the spacetime dimension $d$.    

For odd dimensions $d=2 q+3$, $q=0,1,\ldots$, the Young diagram giving the representation of the Lorentz group is determined by the lengths $\alpha=(m_1,\ldots,m_{q+1})$ of the rows of the diagram, with $0\le m_1\le m_2\le \ldots m_{q+1}$. Let 
$l_r=m_r+(r-1/2)$. Let us call the eigenvalues of a Lorentz transformation $\Lambda$ by $1,\lambda_1=e^{i \epsilon_1},\ldots, \lambda_q=e^{i \epsilon_q}, \lambda_{q+1}=e^{2 \beta}$, and the rest of the eigenvalues are necessarily $\lambda_1^{-1},\ldots,\lambda_{q}^{-1},\lambda_{q+1}^{-1}$. There are three real eigenvalues and the rest are phases (for the Lorentz transformations we are interested in). The character reads 
\bea
\chi_{\alpha}(\Lambda) &=& \frac{A}{B}\,,\nonumber\\  
A &=& \det\{\lambda_i^{l_j}-\lambda_i^{-l_j}\}_{i,j=1,\ldots,q+1}\,,\\
B &=&   \prod_{r=1}^{q+1} (\lambda_{r}^{1/2}-\lambda_{r}^{-1/2})\, \prod_{q+1\ge r> s\ge 1} \left[ (\lambda_{r}+\lambda_{r}^{-1})- (\lambda_{s}+\lambda_{s}^{-1})\right] \nonumber\,.
\eea
For the eigenvalues (\ref{eie}), $A$ and $B$ have in general multiple zeros. Then we expand around the eigenvalues (\ref{eie}) for infinitesimal values of the phases $\epsilon_i$. First we expand the determinant $A$ by the last row
\be
A= \sum_{j=1}^{q+1} (-1)^{j+q+1} (2 \sinh(2 \beta l_j)) \det\left(\{\lambda_i^{l_s}-\lambda_i^{-l_s}\}_{i=1,\ldots,q}^{ s=1,\ldots,\hat{j},\ldots q+1}\right)\,.\label{b5}
\ee
In this last determinant we expand  
\be
\lambda_i^{l_s}-\lambda_i^{-l_s}= \sum_{n \,\,\textrm{odd}} (i \epsilon_i)^n\, 2\frac{l_s^n}{n!}\,. \label{espa}
\ee
Then the determinant in (\ref{b5}) is
\be
\det\left(\left\lbrace\sum_{n \,\,\textrm{odd}} (i \epsilon_i)^n\, 2\frac{l_s^n}{n!}\right\rbrace_{i=1,\ldots,q}^{ s=1,\ldots,\hat{j},\ldots q+1}\right) \,.\label{gg}
\ee
Terms having the same $n$ for two different $s$ do not contribute since they lead to proportional vectors in the index $i$. We have to select in the determinant a different $n$ for each different $s$. To get the lowest order in the expansion we select the values of $n$ to be the first $q$ odd integers. Let $n_0(s)$ be the ordered increasing assignation, and call $\Sigma$ to the permutations of the first $q$ odd integers.   
 We get for the leading term of (\ref{gg}) 
\bea 
2^q\,\det \{(i\epsilon_i)^{2 p-1}\}_{i,p=1,\ldots,q}\,\,\sum_{ \sigma\in \Sigma } (-1)^{\textrm{sig}(\sigma)} \, \prod_{s=1,\ldots,\hat{j},\ldots,q+1} \frac{l_s^{\sigma(n_0(s))}}{\sigma(n_0(s))!} \nonumber \\
=2^q  \det\left(
\left\lbrace\frac{l_s^h}{h!}\right\rbrace_{h=1,3,\ldots,2 q-1} ^{s=1,\ldots,\hat{j},\ldots q+1}\right)\,\, \det\{(i \epsilon_i)^{2 p-1}\}_{i,p=1,\ldots,q}\,.\label{le}
\eea
On the other hand, expanding $B$ we get 
\be
B=2^{q+1}\,\sinh(\beta) (\cosh(2 \beta)-1)^q\, \prod_{r=1}^q (i \epsilon_r) \, \prod_{q\ge r> s \ge 1} ((i \epsilon_r)^2-(i\epsilon_s)^2)\,.
\ee
The leading dependence on $\epsilon_i$ exactly cancels the Vandermonde determinant of powers of $\epsilon_i$ in (\ref{le}). The limit $\epsilon_i\rightarrow 0$ of the character is 
\be
\chi_{\alpha}(\Lambda)= \frac{\sum_{j=1}^{q+1} (-1)^{j+q+1} \sinh(2 \beta l_j) \,\det
\left(\left\lbrace\frac{l_s^h}{h!}\right\rbrace_{h=1,3,\ldots,2 q-1}^{s=1,\ldots,\hat{j},\ldots q+1} \right)}{\sinh(\beta) (\cosh(2 \beta)-1)^q}\,.  
\ee

For even dimensions $d=2 q +2$, $q=1,2,\ldots$, the Young diagram is again determined by the lengths of the rows $\alpha=(m_1,\ldots,m_{q+1})$ in increasing order. Let 
now $l_r=m_r+(r-1)$. The eigenvalues define the $\lambda_i$, $i=1,\ldots,q+1$ as above, but there is no eigenvalue $1$ in the even dimensional case. The character writes
 \bea
\chi_{\alpha}(\Lambda) &=& \frac{A}{B}\,,\nonumber\\  
A &=& \left|\det\{\lambda_i^{l_j}+\lambda_i^{-l_j}\}_{i,j=1,\ldots,q+1}\right|+\left|\det\{\lambda_i^{l_j}-\lambda_i^{-l_j}\}_{i,j=1,\ldots,q+1}\right|\,,\\
B &=&   2 \prod_{q+1\ge r> s\ge 1} \left[ (\lambda_{r}+\lambda_{r}^{-1})- (\lambda_{s}+\lambda_{s}^{-1})\right] \nonumber\,.
\eea
The second determinant in $A$ is subleading in the limit and can be neglected. The first determinant can be expanded as above with the only difference that only even powers appear in the expansion analogous to (\ref{espa}). The result is readily obtained with the same reasoning and is quoted in section \ref{subsec:arbitrary}.

\bibliography{EE}{}
\bibliographystyle{utphys}

\end{document}